\documentclass[12pt, draftclsnofoot, onecolumn]{IEEEtranTCOM}
\usepackage{indentfirst}
\usepackage{multirow}
\usepackage{graphicx}
\usepackage{color}
\usepackage{xcolor}
\usepackage{colortbl} 
\usepackage{listings}
\usepackage{enumerate}
\usepackage{subfigure}
\usepackage{amsmath,amsthm,amscd,amssymb,latexsym,upref,stmaryrd}
\usepackage[noend]{algpseudocode}
\usepackage{algorithmicx,algorithm}
\usepackage{amsfonts,epsfig}
\usepackage{CJK}
\usepackage{float}
\usepackage{cite}
\usepackage{subfigure}
\usepackage{lipsum}
\usepackage{multicol}
\usepackage{setspace}
\usepackage{bm}
\usepackage{amsthm}
\usepackage{amsmath}
\usepackage{amssymb}
\usepackage{amsfonts}
\usepackage{arydshln}
\usepackage{stfloats}
\usepackage{BOONDOX-calo}
\usepackage[aboveskip=-0pt,belowskip=-30pt]{caption}
\usepackage{geometry}
\usepackage{booktabs}
\usepackage{threeparttable}
\newtheorem{remark}{Remark}
\usepackage[utf8]{inputenc}
\usepackage{fancyhdr}
\newtheorem{theorem}{Theorem}
\usepackage[colorlinks,
            linkcolor=purple,
            anchorcolor=purple,
            citecolor=purple]{hyperref}

\renewcommand{\baselinestretch}{1.55}
\begin{document}

\title{Joint Localization and Environment Sensing by Harnessing NLOS Components in RIS-aided mmWave Communication Systems}

\author{Yixuan~Huang, Jie~Yang, Wankai~Tang, Chao-Kai~Wen, Shuqiang~Xia, and Shi~Jin

	\thanks{\setlength{\baselineskip}{14pt}
	Yixuan~Huang, Wankai~Tang, and Shi~Jin are with the National Mobile Communications Research Laboratory, Southeast University, Nanjing, China (e-mail: \{huangyx; tangwk; jinshi\}@seu.edu.cn).
	Jie~Yang is with the School of Automation, Southeast University, Nanjing, China (e-mail: yangjie@seu.edu.cn).
	Jie~Yang, Wankai~Tang, and Shi~Jin are with the Frontiers Science Center for Mobile Information Communication and Security, Southeast University, Nanjing, China.
	Chao-Kai~Wen is with the Institute of Communications Engineering, National Sun Yat-sen University, Kaohsiung, 804, Taiwan (e-mail: chaokai.wen@mail.nsysu.edu.tw).
	Shuqiang~Xia is with the ZTE Corporation and the State Key Laboratory of Mobile Network and Mobile Multimedia Technology, Shenzhen, China (e-mail: xia.shuqiang@zte.com.cn).
}
}

\maketitle	
\vspace{-1cm}
\begin{abstract}

This study explores the use of non-line-of-sight (NLOS) components in millimeter-wave (mmWave) communication systems for joint localization and environment sensing. 
The radar cross section (RCS) of a reconfigurable intelligent surface (RIS) is calculated to develop a general path gain model for RISs and traditional scatterers.
The results show that RISs have a greater potential to assist in localization due to their ability to maintain high RCSs and create strong NLOS links. 
A one-stage linear weighted least squares estimator is proposed to simultaneously determine user equipment (UE) locations, velocities, and scatterer (or RIS) locations using line-of-sight (LOS) and NLOS paths.
The estimator supports environment sensing and UE localization even using only NLOS paths. 
A second-stage estimator is also introduced to improve environment sensing accuracy by considering the nonlinear relationship between UE and scatterer locations. 
Simulation results demonstrate the effectiveness of the proposed estimators in rich scattering environments and the benefits of using NLOS paths for improving UE location accuracy and assisting in environment sensing.
The effects of RIS number, size, and deployment on localization performance are also analyzed.

\end{abstract}

\vspace{0.5cm}

\section{Introduction}

Wireless communication systems offer the potential to realize localization and environment sensing by sharing communication hardware platforms and spectrum resources \cite{tan2021integrated,zhang2022holographic}.
Localization is a crucial aspect for various applications, such as navigation, autonomous vehicle, and extended reality \cite{liu2022survey}. 
In situations where GPS is unavailable, radio-based localization has become a popular solution. 
Millimeter-wave (mmWave) frequency bands, with their large bandwidth, allow for accurate channel parameter estimations, such as time and frequency difference of arrival (TDOA and FDOA) \cite{han2019efficient,guo2017millimeter}.
Massive multiple-input multiple-output (MIMO) antenna arrays, through highly directional transmission, support angle of arrival (AOA) and angle of departure (AOD) estimations \cite{alkhateeb2014channel,zhu2017auxiliary}. 
These parameter estimations are key for localization.

In the past, non-line-of-sight (NLOS) components were often treated as interferences, either being removed \cite{abolfathi2018nlos} or modeled as line-of-sight (LOS) paths with noise \cite{Chen2019improved}.
LOS and NLOS paths can be easily identified due to the large discrepancy in the received signal power \cite{akdeniz2014millimeter}. 
Previous studies on identifying LOS/NLOS paths can be found in \cite{huang2020machine}. 
With improvements in channel estimation techniques \cite{han2019efficient,guo2017millimeter,alkhateeb2014channel,zhu2017auxiliary}, LOS and NLOS path parameters can be effectively extracted. 
NLOS components are valuable for environment sensing, providing information about surrounding environments \cite{liu2021indoor}. 
Some recent studies have used NLOS paths for localization and environment mapping in special reflecting scenarios \cite{yang2022hybrid}. 
Furthermore, NLOS components can improve the localization accuracy of user equipment (UE) \cite{shahmansoori2017position,mendrzik2018harnessing}.

Despite the potential benefits of NLOS paths, their energy loss has posed challenges for practical use, leading to inaccurate channel parameter estimations.
However, this limitation has been overcome with the advent of reconfigurable intelligent surfaces (RISs), which can manipulate electromagnetic waves intelligently \cite{tang2020wireless}. 
Some studies have considered the reflected paths of RISs as LOS paths \cite{yildirim2020modeling}; 
we classify them as NLOS paths \cite{keykhosravi2021semi} due to their similarities with traditional scattered NLOS paths. 
This makes RISs a unique type of scatterer with the advantage of having controllable NLOS paths. 
Adjusting the phase shifts of RIS elements, incident signals onto the RIS array can be aligned in-phase at the receiver, resulting in improved NLOS path gains and accurate channel parameter estimations \cite{tang2021path}. 
This approach overcomes the traditional NLOS path limitations and provides opportunities to enhance the localization and environment sensing performances \cite{zhang2022toward}. 
RISs have been adopted to assist in simultaneous localization and mapping in \cite{yang2022metaslam}.
The Cramér-Rao lower bounds (CRLBs) of the estimated UE location with the help of RISs were studied in literature \cite{elzanaty2021reconfigurable,liu2021reconfigurable}. 
In future communication systems, the deployment of multiple RISs on buildings or mobile vehicles is anticipated, but their locations may be unknown to base stations (BSs) \cite{huang2021transforming}.
Therefore, this study considers the multi-RIS scenario and views RISs as part of the environment.

The similarities and differences between RISs and traditional scatterers have received limited attention in literature. 
\cite{ozdogan2019intelligent} applied traditional electromagnetic analysis methods to compare the path loss models of scatterers and RISs. 
Measurements of received signal power show good agreements between RISs and metal plates in their reflecting abilities \cite{tang2021path}. 
In the case of NLOS paths, the radar cross section (RCS) plays a crucial role in path loss \cite{goldsmith2005wireless}. 
RCS is defined as the ratio of scattered signal energy to incident signal energy \cite{Merrill1990radar} and depends on wavelength, incident, and scattering angles; however, it is independent of waveform and the distance between scatterers and observers \cite{balanis2012advanced}.
In previous localization-related works, the RCS of scatterers was either ignored, modeled as a random variable \cite{yan2018robust}, or given by a determined value \cite{wymeersch2020beyond}.
Moreover, the RCS of RISs was rarely discussed in literature, with only simple models presented in \cite{basar2020simris} and \cite{wang2021received}. 
This study derives the RCS of RISs based on the popular path loss model given in \cite{tang2021path} under the far-field assumption.
It also sheds light on the similarities and differences between the scattering characteristics of scatterers and RISs, including the consistency between their path loss models.

With the integration of RISs, we aim to improve localization and environment sensing capabilities.
Literature have extensively studied various localization estimators that use channel parameters as functions of unknown location parameters, such as linear least squares \cite{van2015least}, linear weighted least squares (WLS) \cite{yang2021model}, and belief propagation \cite{yu2021efficient}. 
We adopt the WLS estimator because of its balance of estimation accuracy and computational complexity \cite{yang2021model}. 
In the WLS estimator development, the use of a single type of measurement \cite{amiri2017asymptotically,wang2015asymptotically} has evolved to the integration of delay, Doppler, and angle measurements \cite{amiri2017efficient,ho2004accurate}. 
Specifically, the best UE location and velocity estimation accuracy is achieved by combining TDOA/FDOA/AOA measurements in a one-stage WLS estimator \cite{yang2021model}.
The effects of NLOS paths on UE localization have been analyzed in \cite{shahmansoori2017position,mendrzik2018harnessing}; however, the utilization of NLOS paths with WLS estimators for UE localization is still rare. 
In addition, localization and environment sensing are typically studied independently, where NLOS paths are used to estimate scatterer locations and velocities with known UE locations in \cite{yang2021model} or to locate the UE with estimated scatterer locations in \cite{chen2016elliptical}. 
Using the WLS estimator, we propose a one-stage closed-form solution for the simultaneous estimation of UE locations, velocities, and scatterer locations by integrating the TDOA/FDOA/AOA measurements of LOS paths and AOA/AOD measurements of NLOS paths. 
This solution surpasses the state-of-the-art method in \cite{yang2021model} and can locate UE using only NLOS paths. 
A second-stage estimator is also proposed to further improve environment sensing accuracy.

\clearpage
Our key contributions are summarized as follows:

\begin{itemize}
\item \textbf{General model for RISs and scatterers.} Comparison of RISs and traditional scatterers, derivation of a general path gain model, and proof that RISs can provide strong NLOS paths and high-quality environmental information.

\item \textbf{Joint 3D localization and environment sensing.} Proposal of a one-stage estimator for the simultaneous estimation of UE locations, velocities, and scatterer locations, surpassing previous methods and approaching CRLBs under low noise levels, and a second-stage estimator that further enhances environment sensing performance.
\end{itemize}

{\bf Notations}---The scalars (e.g., $a$) are denoted in italic, vectors (e.g., $\mathbf{a}$) in bold, and matrices (e.g., $\mathbf{A}$) in bold capital letters. 
The true values of estimated variables $a$, $\mathbf{a}$, and $\mathbf{A}$ are given by $a^\circ$, $\mathbf{a}^\circ$, and $\mathbf{A}^\circ$, respectively.
$\mathbf{a}(i)$ represents the $i$-th element in vector $\mathbf{a}$, whose $\ell_{2}$-norm is $\|\mathbf{a}\|$.
$\mathbf{A}_{i,j}$ is the element of the $i$-th row and $j$-th column in matrix $\mathbf{A}$; $\mathbf{A}^{-1}$ is the inverse matrix.
$j = \sqrt{-1}$ is the imaginary unit. 
$\Re(\cdot)$ takes the real component of a complex number, whereas $|\cdot|$ denotes the module of a complex value.
Transpose and Hermitian operators are presented as $(\cdot)^T$ and $(\cdot)^H$, respectively. 
$\text{diag}\{\cdot\}$ and $\text{blkdiag}(\mathbf{A}_1,\ldots,\mathbf{A}_k)$ denote diagonal and block diagonal matrices, respectively.
$\mathbb{E}\{\cdot\}$ denotes statistical expectation, and the Kronecker product is presented by $\otimes$.

\begin{figure}
	\centering
	\includegraphics[width=0.65\textwidth]{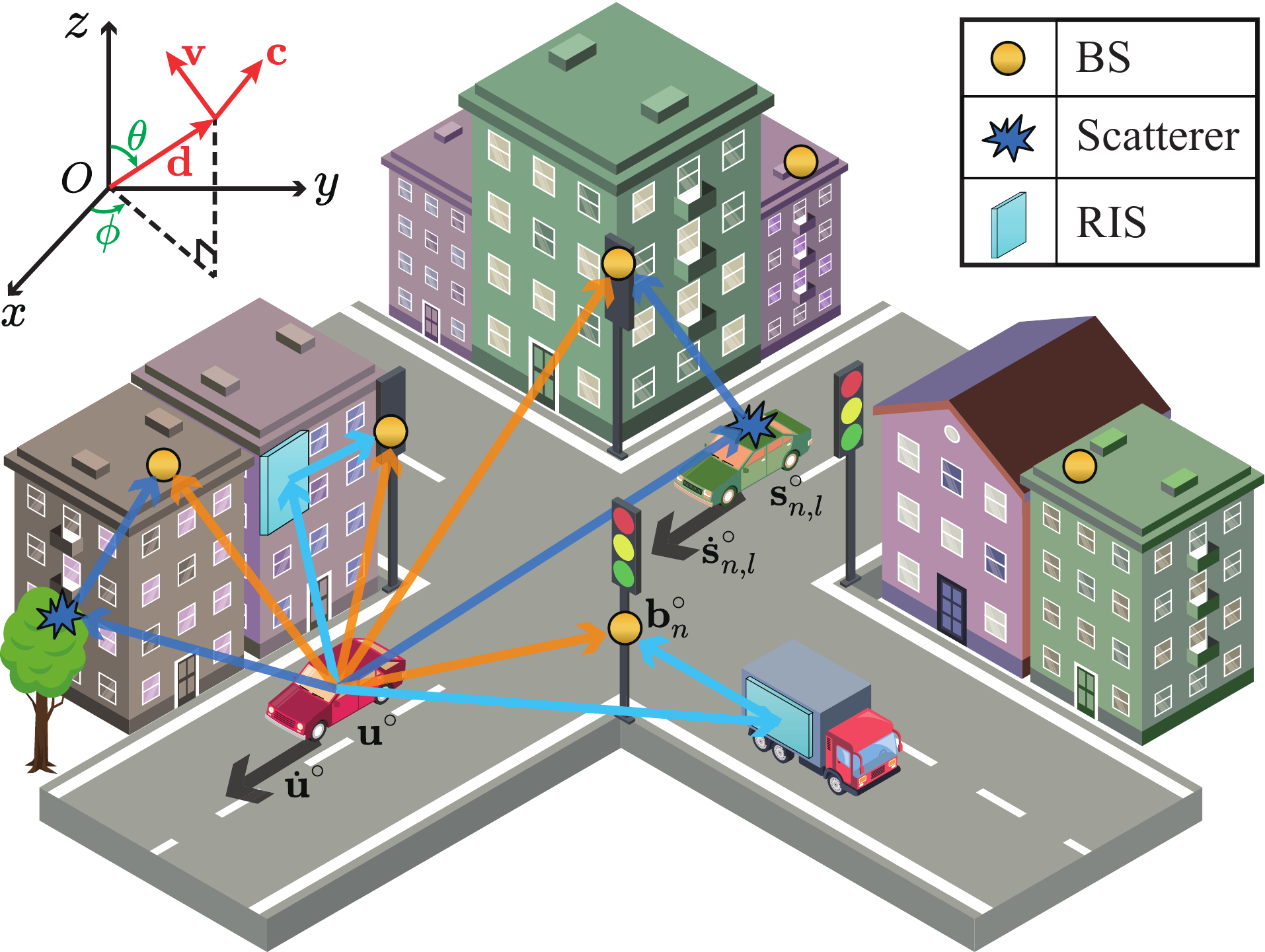}	
	\captionsetup{font=footnotesize}
	\caption{Illustration of an RIS-aided mmWave communication system with densely deployed BSs.}
	\label{fig:model}
\end{figure}

\begin{table}
	\vspace{0.7cm}
	\centering
	\caption{Notations of important variables.}\label{tab:notation}
	\vspace{0.3cm}
	\renewcommand{\arraystretch}{1.78}
	\fontsize{8}{8}\selectfont
	\begin{tabular}{m{1.25cm} m{6.25cm} m{0.9cm} m{6.8cm}}
		\hline		
		Notation & Definition & Notation & Definition \\
		\hline
		$f$ & center frequency & $\lambda$ & wavelength\\
		\hline
		$(\mathbf{d}, \mathbf{c}, \mathbf{v})$ & an orthonormal basis in the 3-D space & $g_{n,l}^\circ$ & complex gain of the $(n,l)$-th path\\
		\hline
		$N$ & number of BSs & $\tau_{n,l}^\circ, \nu_{n,l}^\circ$ & delay and Doppler shift of the $(n,l)$-th path, respectively\\
		\hline
		$L_n$ & number of scatterers or RISs linked to the $n$-th BS & $d_{n,0}$ & distance between the $n$-th BS and UE \\
		\hline
		$N_{\rm{s}}$ & $N_{\rm{s}} = \sum_{n=1}^N L_n$, total number of scatterers or RISs & $d_{n,l}^{\star}$ & distance from the $(n,l)$-th scatterer to BS or UE, $\star \!\!\in\!\! \{\rm{r},\! \rm{t}\}$ \\
		\hline
		$\mathbf{b}_n$ & location of the $n$-th BS & $\kappa$ & $\kappa = 2\pi/\lambda$, wave number \\
		\hline
		$\mathbf{u}^\circ,{\dot{\mathbf{u}}^\circ}$ & location and velocity of the UE, respectively & $A$ & reflecting coefficient of RIS elements \\
		\hline
		$\mathbf{s}_{n,l}^\circ,\dot{\mathbf{s}}_{n,l}^{\circ}$ & location and velocity of the $(n,l)$-th scatterer or RIS & $\xi_{k,m}$ & tunable phase shift of the $(k,m)$-th RIS element \\
		\hline
		$r_{n1,l}^\circ,\dot{r}_{n1,l}^\circ$ & TDOA/FDOA-related parameter of the $(n,l)$-th path & $G$ & antenna gain of transmitting and receiving antennas \\
		\hline
		$(\phi_{n,l}^{\rm{r}\circ},\theta_{n,l}^{\rm{r}\circ})$ & azimuth and elevation AOA of the $(n,l)$-th path & $G_{\rm{RIS}}$ & scattering gain of RIS elements \\
		\hline
		$(\phi_{n,l}^{\rm{t}\circ},\theta_{n,l}^{\rm{t}\circ})$ & azimuth and elevation AOD of the $(n,l)$-th path & $\mathbf{p}_{n,l,m_\star}$ & position of the $m_\star \!$-th antenna of the $(n,l)$-th path, $\star \!\!\in\!\! \{\rm{r},\! \rm{t}\}$ \\
		\hline
		$S$ & total area of the PEC scatterer or RIS & $\mathbf{p}_{k,m}$ & position of the $(k,m)$-th scatterer subpatch or RIS element\\
		\hline
		$S_{\rm{sub}}$ & area of a scatterer subpatch or RIS element & $\alpha_{k,m}$ & phase shift of the path reflected by the $(k,m)$-th subpatch\\
		\hline
		$\sigma$ & RCS of the scatterer or RIS  & $\mathbf{m}^\circ$ & noise-free parameters used in the first-stage estimator\\
		\hline
		$\sigma^{\rm{subpatch}}_{k,m}$ & RCS of the $(k,m)$-th scatterer subpatch  & $\mathbf{m}_{n,l}^\circ$ & noise-free parameters used in the second-stage estimator \\
		\hline
		$\sigma_{\rm{subpatch}}$ & RCS of the scatterer subpatch in the far field & $\sigma_{\rm{subRIS}}$ & RCS of the RIS element in the far field\\
		\hline
		$\mathcal{d}_{k,m}^{\star}$ & distance from the center of the $(k,m)$-th scatterer & $\mathbf{s}_{n,l}^{\rm{e}1}$ & rough estimate of the $(n,l)$-th scatterer location \\
		& subpatch or RIS element to BS or UE, $\star \!\!\in\!\! \{\rm{r},\! \rm{t}\}$ & & in the first-stage estimator\\
		\hline
		$(\!\varphi_{k,m}^{\star}\!,\vartheta_{k,m}^{\star}\!)$ & azimuth and elevation angles from the $(k,m)$-th scat- & $(\varphi_{\star},\vartheta_{\star})$ & azimuth and elevation angles from the center of \\
		& terer subpatch or RIS element to BS or UE, $\star \!\!\in\!\! \{\rm{r},\! \rm{t}\}$ & & PEC scatterer or RIS to BS or UE, $\star \!\!\in\!\! \{\rm{r},\! \rm{t}\}$\\
		\hline
	\end{tabular}
	\vspace{-0.8cm}
\end{table}

\section{System Model}
\label{sec:system-model}

We consider an RIS-aided mmWave communication system, as shown in Fig. \ref{fig:model}. 
It has $N$ BSs with clocks synchronized. Center frequency is denoted as $f$, and the wavelength is $\lambda$. 
The system uses orthogonal frequency division multiplexing (OFDM) symbols for communication.
BSs and UE employ uniform planar arrays (UPAs) with $M_{\rm{r}}$ and $M_{\rm{t}}$ antenna elements, respectively.
For simplicity, we only focus on a single UE scenario in this study. 
However, the results can be applied to multiple UE conditions, provided that the pilot signals for each UE are orthogonal, and any multi-user interference (MUI) contributes to the additive noise in the received signals. 
The key variables are summarized in Table \ref{tab:notation}.

\subsection{System Geometry}

The system works in the 3D space $\mathbb{R}^{3}=\left\{[x, y, z]^{T}: x, y, z \in \mathbb{R}\right\}$. 
The location of the $n$-th BS is $\mathbf{b}_n=[x^{\rm{b}}_n,y^{\rm{b}}_n,z^{\rm{b}}_n]^T$, for $ n=1, 2, \ldots, N$. 
$\mathbf{u}^\circ=[x^\circ,y^\circ,z^\circ]^T$ and $\dot{\mathbf{u}}^\circ=[\dot{x}^\circ,\dot{y}^\circ,\dot{z}^\circ]^T$ represent the true values of unknown UE location and velocity, whereas the estimated counterparts are given by $\mathbf{u}$ and $\dot{\mathbf{u}}$, respectively.
Given the sparsity and high path loss of the mmWave channel \cite{akdeniz2014millimeter}, our focus is on LOS paths and single-bounce NLOS paths.
Multi-bounce signals are treated as background noise.
The location of the $l$-th scatterer\footnote{
When a large target (e.g., buildings) is illuminated by the uplink signals, multiple single-bounce NLOS paths are established between UE and BSs.
Each NLOS path corresponds to one scatterer position, as long as the paths are resolvable in temporal or angular domain \cite{mendrzik2018harnessing}.
}
between the $n$-th BS and UE is denoted as $\mathbf{s}^\circ_{n,l}=[x^{{\rm{s}}\circ}_{n,l}, y^{{\rm{s}}\circ}_{n,l}, z^{{\rm{s}}\circ}_{n,l}]^T$, whereas the velocity is $\dot{\mathbf{s}}^\circ_{n,l}=[\dot{x}^{{\rm{s}}\circ}_{n,l}, \dot{y}^{{\rm{s}}\circ}_{n,l}, \dot{z}^{{\rm{s}}\circ}_{n,l}]^T$ for $l=1, 2, \ldots, L_n$. 
$L_n$ is the number of scatterers between the $n$-th BS and UE. $N_{\rm{s}} = \sum_{n=1}^N L_n$ denotes the total number of scatterers. 
Taking RISs as a unique type of scatterer, the location of the $(n, l)$-th RIS is also denoted as $\mathbf{s}^\circ_{n,l}$.
The BSs and UE are assumed to be in the far fields of scatterers and RISs, meaning that the distance between them should be larger than $2D^2/\lambda$, where $D$ is the largest dimension of scatterers or RISs \cite{tang2020wireless}. 
Notably, $\partial \mathbf{u}^\circ/ \partial t=\dot{\mathbf{u}}^\circ$, and $\partial \mathbf{s}^\circ_{n,l}/\partial t=\dot{\mathbf{s}}^\circ_{n,l}$.
The aim of our work is to estimate $\mathbf{u}^{\circ}$, $\dot{\mathbf{u}}^{\circ}$, $\mathbf{s}_{n,l}^{\circ}$, and $\dot{\mathbf{s}}_{n,l}^{\circ}$ by using the signals received at BSs.

\subsection{Multipath Channel Model}

We assume that $L_n+1$ multipath components exist between the $n$-th BS and UE, with one LOS path and $L_n$ NLOS paths.
The RIS-reflected path is regarded as a unique type of NLOS path, given the similar geometric relationships between them.
We focus on the uplink channel because channel estimation and localization are assumed to be accomplished by BSs.
The channel impulse response of the $n$-th BS is variable in time and frequency, given in the matrix form as \cite{heath2016overview}
\vspace{-0.1cm}
\begin{equation}
\label{eq:channel-model-matrix}
\mathbf{H}_n(t, f) = \sum_{l=0}^{L_{n}}g_{n,l}^\circ e^{j2\pi \left(\nu_{n,l}^\circ t - \tau_{n,l}^\circ f\right)} \mathbf{a}_{\rm{r}}(\phi^{{\rm{r}}\circ}_{n, l}, \theta^{{\rm{r}}\circ}_{n, l}) \mathbf{a}_{\rm{t}}^H(\phi^{{\rm{t}}\circ}_{n, l}, \theta^{{\rm{t}}\circ}_{n, l}),
\vspace{-0.1cm}
\end{equation}
where $l\!=\!0$ represents the LOS path, and $l\! >\! 0$ represents the NLOS path; $g^\circ_{n,l}$ is the complex path gain of the $(n,l)$-th path, where the $(n,l)$-th path represents the $l$-th path between the $n$-th BS and UE; 
$\tau^\circ_{n,l}$ and $\nu^\circ_{n,l}$ represent the delay and Doppler shift, respectively; 
$\mathbf{a}_{\rm{r}}(\phi^{{\rm{r}}\circ}_{n, l}, \theta^{{\rm{r}}\circ}_{n, l})$ and $\mathbf{a}_{\rm{t}}(\phi^{{\rm{t}}\circ}_{n, l}, \theta^{{\rm{t}}\circ}_{n, l})$ are the steering vectors of the UPA of the $n$-th BS and UE, respectively; the azimuth and elevation angles are denoted by $\phi_{n, l}^{\star\circ}$ and $\theta_{n, l}^{\star\circ}$, where $\star \!\in \!\{\rm{r}, \rm{t}\}$.
As illustrated in Fig. \ref{fig:model}, we define an orthonormal basis $(\mathbf{d}, \mathbf{c}, \mathbf{v})$ in the 3D space with the azimuth and elevation angles $(\phi, \theta)$, given as
{\setlength\abovedisplayskip{4pt}
\setlength\belowdisplayskip{4pt}
\begin{align}
\mathbf{d}\left(\phi, \theta\right)&=\left[\cos (\phi) \sin (\theta), \sin (\phi) \sin (\theta), \cos (\theta)\right]^{T}, \\[-1.5mm]
\mathbf{c}\left(\phi, \theta\right)&=\left[-\sin (\phi), \cos (\phi), 0\right]^{T}, \\[-1.5mm]
\mathbf{v}\left(\phi, \theta\right)&=\left[-\cos (\phi) \cos (\theta),-\sin (\phi) \cos (\theta), \sin (\theta)\right]^{T},
\end{align}}
\hspace{-0.15cm}where vector $\mathbf{d}$ is used to construct the steering vectors of UPAs, and vectors $\mathbf{c}$ and $\mathbf{v}$ are employed to establish the relationships between the channel and location parameters in Sec. \ref{sec:wls}.
We define the position of the $m_{\star}$-th transmitting/receiving antenna of the $(n,l)$-th path as $\mathbf{p}_{n,l,m_{\star}} \!\!=\! \left[x_{n,l,m_{\star}}\!, y_{n,l,m_{\star}}\!, z_{n,l,m_{\star}}\right]^T$.
Taking the origin as the reference point, the steering vector $\mathbf{a}_{\star}(\phi^{\star\circ}_{n, l}, \theta^{\star\circ}_{n, l})$ is given as
\begin{equation}
\mathbf{a}_{\star}(\phi^{\star\circ}_{n, l}, \theta^{\star\circ}_{n, l}) = e^{-j \frac{2 \pi}{\lambda} \mathbf{P}_{n, l}^\star \mathbf{d}(\phi^{\star\circ}_{n, l}, \theta^{\star\circ}_{n, l})},
\vspace{-0.15cm}
\end{equation}
where $\mathbf{P}_{n,l}^\star = \left[\mathbf{p}_{n,l,1}, \mathbf{p}_{n,l,2}, \ldots, \mathbf{p}_{n,l,M_{\star}}\right]^T$ is the matrix composed of antenna positions.

Next, we model the complex path gains of LOS and NLOS paths.
\begin{itemize}
\item \textbf{LOS path}:
The LOS path gain between the UE and the $n$-th BS is given as \cite{goldsmith2005wireless}
\begin{equation}
g_{n,0}^\circ = \frac{\lambda \sqrt{G} e^{-\frac{j 2 \pi d_{n,0}}{\lambda}}}{4 \pi d_{n,0}},
\end{equation}
where $G$ is the integrated antenna gain of transmitting and receiving antennas; $d_{n,0}$ represents the distance between the $n$-th BS and UE. 

\item \textbf{NLOS path of traditional scatterer}:
The gain of the $(n,l)$-th scattered path is given as \cite{goldsmith2005wireless}
\begin{equation}
\label{eq:nlos-pathgain-scatterer}
g_{n,l}^{\circ}=\frac{\lambda \sqrt{G \sigma} e^{-\frac{j 2 \pi (d_{n,l}^{\rm{r}} + d_{n,l}^{\rm{t}})}{\lambda}}}{(4 \pi)^{\frac{3}{2}} d_{n,l}^{\rm{r}} d_{n,l}^{\rm{t}}},
\end{equation}
where $\sigma$ denotes the RCS of the scatterer; $d_{n,l}^{\rm{r}}$ and $d_{n,l}^{\rm{t}}$ are the distances from the center of the $(n,l)$-th scatterer to the $n$-th BS and UE, respectively, as displayed in Fig. \ref{fig:structure-scatterer-ris}(a). 

\item \textbf{NLOS path of RIS}:
Assuming that the RIS is composed of $K \times M$ tunable elements, the complex gain of the $(n,l)$-th RIS-aided path is given as \cite{tang2021path}
\begin{equation}
\label{eq:nlos-pathgain-ris}
g^{\circ}_{n,l} = \lambda  \sqrt{\frac{G G_{{\rm{R I S}}}  d_{{\rm{x}}} d_{{\rm{y}}}}{(4 \pi)^3}} \times \sum_{k=1}^{K} \sum_{m=1}^{M}  \frac{\sqrt{F(\varphi_{k,m}^{\rm{r}}, \vartheta_{k,m}^{\rm{r}}) F(\varphi_{k,m}^{\rm{t}}, \vartheta_{k,m}^{\rm{t}})} A e^{-j \xi_{k, m}}}{\mathcal{d}_{k, m}^{{\rm{r}}}\mathcal{d}_{k, m}^{{\rm{t}}}}e^{\frac{-j 2 \pi\left(\mathcal{d}_{k, m}^{{\rm{r}}}+\mathcal{d}_{k, m}^{{\rm{t}}}\right)}{\lambda}},
\end{equation}
where $G_{\rm{RIS}} $ is the scattering gain of RIS elements; 
$\left(\varphi_{k,m}^{{\rm{r}}}, \vartheta_{k,m}^{{\rm{r}}}\right)$ and $\left(\varphi_{k,m}^{{\rm{t}}}, \vartheta_{k,m}^{{\rm{t}}}\right)$ denote the azimuth and elevation angles pointing from the center of the $(k,m)$-th RIS element to BS and UE, and the corresponding distances are $\mathcal{d}_{k,m}^{\rm{r}}$ and $\mathcal{d}_{k,m}^{\rm{t}}$, respectively; 
$d_{\rm{x}}$ and $d_{\rm{y}}$ represent the size of a single RIS element, as shown in Fig. \ref{fig:structure-scatterer-ris}(b).
$A$ and $\xi_{k,m}$ represent the reflecting coefficient and tunable phase shift of the $(k,m)$-th RIS element, respectively. 
Here, the $(k,m)$-th RIS element means the element located at the $k$-th row and the $m$-th column in the RIS array.
$F\left(\varphi_{k,m}^{{\rm{r}}},\vartheta_{k,m}^{{\rm{r}}}\right)$ and $F\left(\varphi_{k,m}^{{\rm{t}}},\vartheta_{k,m}^{{\rm{t}}}\right) $ are the normalized power radiation patterns of the RIS element in the directions of reflecting and receiving with an example as \cite{stutzman2012antenna}
\begin{equation}
\label{eq:pattern}
F(\varphi, \vartheta)=\left\{\begin{aligned}
&\cos^\beta \vartheta, & &\vartheta \in\left[0, \frac{\pi}{2}\right], &&\varphi \in[0,2 \pi], \\[-2pt]
&0, & &\vartheta \in\left(\frac{\pi}{2}, \pi\right],&& \varphi \in[0,2 \pi],
\end{aligned}\right.
\end{equation}
where $\beta$ represents the scattering directivity of RIS elements.

\end{itemize}

\subsection{Relationships between Channel and Location Parameters}
\label{sec:relation-parameter}
Joint localization and environment sensing are executed with estimated channel parameters. 
We bridge the channel and location parameters in this subsection.

\begin{itemize}
\item \textbf{TDOA}: Without loss of generality, we take the time of arrival of the LOS path of the first BS as the reference time. The TDOA-related parameter $r_{n1,l}^{\circ}$ for the $(n,l)$-th path is defined by
\vspace{-0.15cm}
\begin{equation}\label{eq:tdoa}
r_{n1,l}^{\circ} = v(\tau ^\circ_{n,l}-\tau ^\circ_{1,0}) = r_{n,l}^{\circ}-r_{1,0}^\circ,
\vspace{-0.15cm}
\end{equation}
based on the geometric information involved in the delay
\begin{equation}\label{eq:tdoa-goe}
r_{n,l}^{\circ} = v(\tau ^\circ_{n,l}-\omega ) = \left\{\begin{aligned}
&\left \| \mathbf{u}^\circ -\mathbf{b}_n \right \|, &&l=0, \\[-6pt]
&\left \| \mathbf{u}^\circ - \mathbf{s}^\circ_{n,l}\right \|  + \left \|  \mathbf{s}^\circ_{n,l} - \mathbf{b}_n\right \|,&& l>0,
\end{aligned}\right.
\end{equation}
where $v$ is the propagation velocity of the signals; $r_{n,l}^\circ$ denotes the propagation distance of the $(n,l)$-th path; $\omega$ is the unknown clock bias between the BSs and UE. 

\item \textbf{FDOA}: Similarly, taking the Doppler shift of the LOS path of the first BS as the reference, the FDOA-related parameter $\dot{r}^{\circ}_{n1,l}$ for the $(n,l)$-th path is defined by
\vspace{-0.15cm}
\begin{equation}\label{eq:fdoa}
\dot{r}^{\circ}_{n1,l}  = \lambda(\nu^\circ_{n,l} - \nu^\circ_{1,0}) = \dot{r}^{\circ}_{n,l}-\dot{r}^{\circ}_{1,0},
\vspace{-0.15cm}
\end{equation}
based on
\begin{equation}\label{eq:fdoa-goe}
\dot{r}_{n,l}^{\circ} = \frac{\partial r^\circ_{n,l}}{\partial t}  = \left\{\begin{aligned}
&\frac{\dot{\mathbf{{u}}}^{\circ {T}} ({\mathbf{{u}}}^{\circ} - \mathbf{{b} }_n)}{\left \| \mathbf{{u} }^\circ  - \mathbf{{b} }_n \right \| }, &&l=0, \\
&\frac{(\dot{\mathbf{{u} }}^\circ -  \dot{\mathbf{{s} }}^\circ_{n,l})^{T}({\mathbf{{u} }^\circ} - {\mathbf{{s} }^\circ_{n,l}})}{\left \| {\mathbf{{u} }^\circ} - {\mathbf{{s} }^\circ_{n,l}} \right \| } + \frac{\dot{\mathbf{{s}}}^{\circ{T}}_{n,l}-({\mathbf{{s} }^\circ_{n,l}}-\mathbf{{b}}_n)}{\left \| {\mathbf{{s} }^\circ_{n,l}}-\mathbf{{b}}_n \right \| },&& l>0,
\end{aligned}\right.
\end{equation}
where $\nu_{n,l}^\circ$ is the Doppler shift of the $(n,l)$-th path; $\dot{r}^\circ_{n,l}$ is the varying rate of the propagation distance of the $(n,l)$-th path. 

\item \textbf{AOA/AOD}: For angular channel parameters, the AOA of the LOS path is defined as
\begin{equation}\label{eq:los-aoa}
\phi^{{\rm{r}}\circ}_{n,0} = \arctan\frac{y^{\circ} - y^{\rm{b}}_n}{x^{\circ} - x^{\rm{b}}_n} ,\quad \theta^{{\rm{r}}\circ}_{n,0} = \arccos\frac{z^{\circ} - z^{\rm{b}}_n}{\left \| \mathbf{{u}}^\circ - \mathbf{{b}}_n \right \| },
\end{equation}
whereas the AOD of the LOS path does not provide extra information and is not used in this study.
For the NLOS path of scatterer or RIS, the AOA is given as
\begin{equation}\label{eq:nlos-aoa}
\phi^{{\rm{r}}\circ}_{n,l} = \arctan\frac{y^{{\rm{s}}\circ}_{n,l} - y^{\rm{b}}_n}{x^{{\rm{s}}\circ}_{n,l} - x^{\rm{b}}_n} ,\quad \theta^{{\rm{r}}\circ}_{n,l} = \arccos\frac{z^{{\rm{s}}\circ}_{n,l} - z^{\rm{b}}_n}{\left \| \mathbf{{s}}^\circ_{n,l} - \mathbf{{b}}_n \right \| },
\end{equation}
and the AOD as
\begin{equation}\label{eq:nlos-aod}
\phi_{n, l}^{{\rm{t}}\circ}=\arctan \frac{y^{\circ}-y_{n, l}^{{\rm{s}}{\circ}}}{x^{\circ}-x_{n, l}^{{\rm{s}}{\circ}}}, \quad \theta_{n, l}^{{\rm{t}}\circ}=\arccos \frac{z^{\circ}-z_{n, l}^{{\rm{s}} \circ}}{\left\|\mathbf{u}^{\circ}-\mathbf{s}_{n, l}^{\circ}\right\|}.
\end{equation}
\end{itemize}

To summarize, the relationships between the TDOA/FDOA/AOA/AOD-related channel and location parameters are characterized by Equations \eqref{eq:tdoa}-\eqref{eq:nlos-aod}.
In the next section, we first evaluate the scattering properties of RISs in comparison to traditional scatterers, demonstrating that RISs can enhance the RCS of NLOS paths. Subsequently, in Sec. \ref{sec:wls}, the channel parameters of NLOS paths are employed to assist in localization and environment sensing.

\section{RCS Comparisons between Scatterers and RISs}
\label{sec:rcs}

Although NLOS paths are proven to contain certain information about surrounding environments \cite{liu2021indoor,yang2022hybrid} and assist in UE localization \cite{shahmansoori2017position,mendrzik2018harnessing}, large pathloss in mmWave frequency bands prevents them from practical use.
By contrast, the use of RISs with tunable phase shifts can enhance NLOS paths and improve localization accuracy.
In this section, we first derive the RCS of RISs and compare it with that of traditional scatterers. 
We aim to reveal the similarities and differences between their scattering properties, providing evidence for the use of RISs in localization.
For a fair comparison, the traditional scatterer used in this study is a rectangular perfectly electrical conductor (PEC) plate of the same size and shape as RISs and with negligible thickness.

\subsection{RCS of PEC Scatterer}

The formal definition of the RCS is given by \cite{Merrill1990radar}
\begin{equation}
\label{eq:rcs-definition}
\sigma=\lim _{d \rightarrow \infty} 4 \pi d^{2} \frac{\left|{E}_{{\rm{r}}}\right|^{2}}{\left|{E}_{\rm{in}}\right|^{2}},
\end{equation}
where ${E}_{\rm{in}}$ and ${E}_{{\rm{r}}}$ are the electric field intensity of the incident wave impinging on the object and that of the scattered wave observed at the receiver, respectively; $d$ represents the distance between the scatterer and the receiver. 
Bistatic RCS is used in this study because the transmitter and receiver are usually not located at the same position in communication systems.

\begin{figure}
\centering
\captionsetup{font=footnotesize}
  \hspace{-0.5cm}
  \begin{minipage}{0.49\linewidth}
    \centerline{\quad\quad\quad \includegraphics[width=0.7\textwidth]{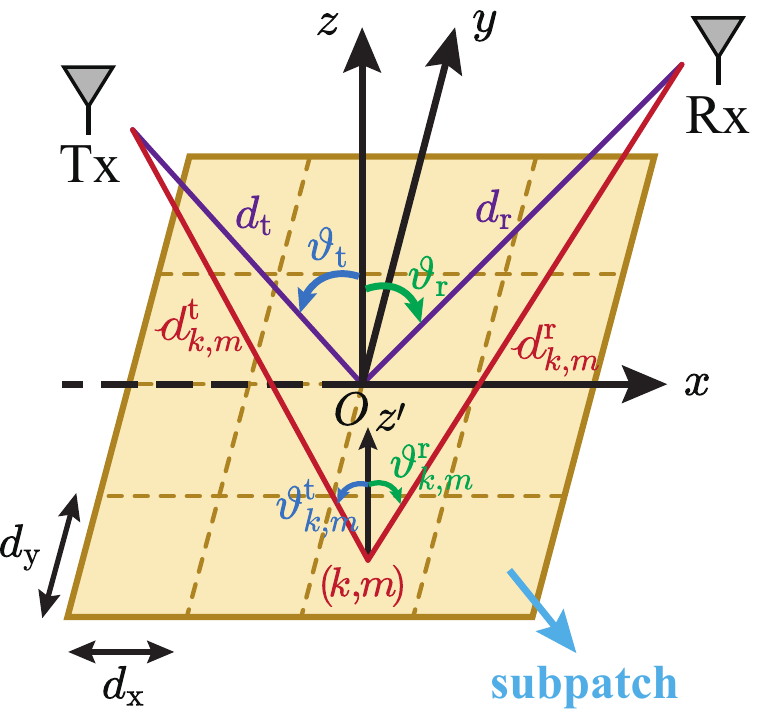}}
    \vspace{-0.2cm}
    \centerline{\footnotesize{(a) Scatterer}}
  \end{minipage}
  \hfill
  \begin{minipage}{0.49\linewidth}
    \centerline{\quad\quad\quad \includegraphics[width=0.7\textwidth]{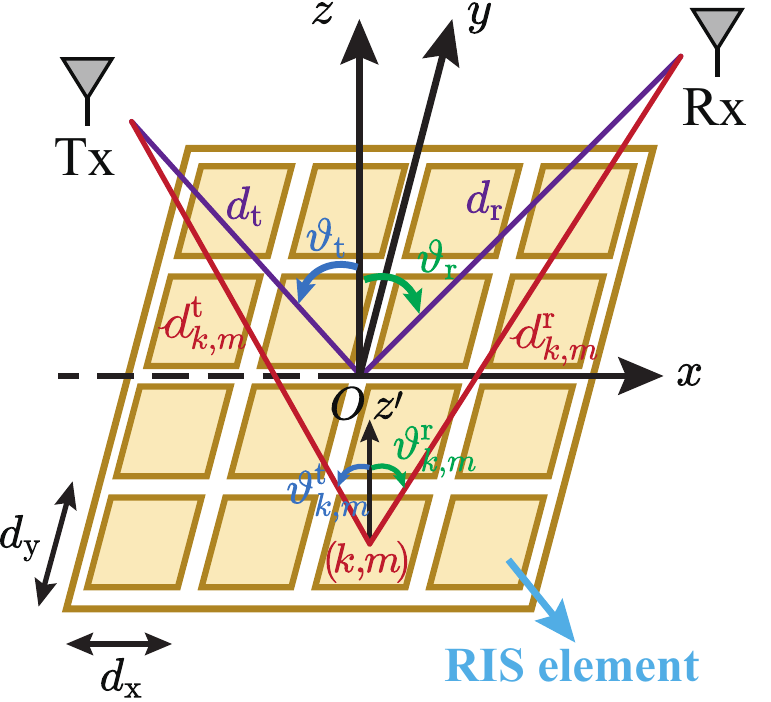}}
    \vspace{-0.2cm}
    \centerline{\footnotesize{(b) RIS}}
  \end{minipage}
  \vspace{0.2cm}
  \caption{Comparison of the physical structures between a scatterer and an RIS ($d_{n,l}^{\rm{r}}$ and $d_{n,l}^{\rm{t}}$ are simplified as $d_{\rm{r}}$ and $d_{\rm{t}}$, respectively).}
  \label{fig:structure-scatterer-ris}
  \vspace{0.2cm}
\end{figure}

Fig. \ref{fig:structure-scatterer-ris} presents a comparison of the physical structures of a scatterer and an RIS. 
Suppose that the RIS consists of $K\times M$ RIS elements and the area of each element $S_{\rm{sub}} = d_{\rm{x}} \times d_{\rm{y}}$.
Then, the total area of the RIS $S=KMd_{\rm{x}}d_{\rm{y}}$. 
Fair comparisons between the RIS and scatterer require that the area of the rectangular PEC plate is equal to that of the RIS. 
Therefore, we divide the rectangular scatterer into $K\times M$ subpatches as an analogy to RIS, and the area of each subpatch is $S_{\rm{sub}}$.
Then, the RCS of the scatterer can be calculated by summing up the RCSs of all subpatches while considering the phase shifts caused by interelement path length differences, given by \cite{Merrill1990radar}
\begin{equation}
\label{eq:rcs-scatterer-summation-subpatch}
\sigma_{\rm{s}} = \left|\sum_{k = 1}^K\sum_{m = 1}^M\sqrt{\sigma_{k,m}^{\rm{subpatch}}}e^{\frac{-j 2 \pi\left(\mathcal{d}_{k, m}^{\rm{r}}+\mathcal{d}_{k, m}^{\rm{t}}\right)}{\lambda}}\right|^2,
\end{equation}
where $\sigma_{k,m}^{\rm{subpatch}}$ is the RCS of the $(k,m)$-th subpatch.
Among the methods of RCS calculation \cite{Merrill1990radar,balanis2012advanced}, we choose the method of physical optics (PO) 
to obtain $\sigma_{k,m}^{\rm{subpatch}}$ due to its concise expression, polarization independence, and high accuracy for small incident angles, given by\footnote{The method of PO is initially derived for the RCS of the whole scatterer by surface integral, which can be exactly illustrated by dividing the scatterer into plenty of small subpatches, given in the discrete form as \eqref{eq:rcs-scatterer-summation-subpatch}. Simulation results showed that the RCSs calculated by the superposition of subpatch RCSs as \eqref{eq:rcs-scatterer-summation-subpatch} and by direct PO method as \eqref{eq:rcs-scatterer-po-bis} fit well with each other \cite{Merrill1990radar}.}
\begin{equation}
\label{eq:rcs-scatterer-po-bis}
\sigma_{k,m}^{\rm{subpatch}}= \frac{4\pi S_{\rm{sub}}^2}{\lambda^2} \cos ^2\left(\vartheta^{{\rm{t}}}_{k,m}\right) \text{Sa}^2\left( \frac{\kappa d_{\rm{x}}}{2}\left(\sin \left(\vartheta^{{\rm{r}}}_{k,m}\right)+\sin \left(\vartheta^{{\rm{t}}}_{k,m}\right)\right)\right),
\end{equation}
where $\text{Sa}(x) = \sin x/x$, and $\text{Sa}(0)=1 $; $\kappa=2\pi/\lambda$ is the wave number; $\vartheta^{\rm{r}}_{k,m}$ and $\vartheta^{\rm{t}}_{k,m}$ denote the elevation angles\footnote{Considering that the normalized power radiation pattern of RIS elements given in \eqref{eq:pattern} only considers the elevation angles, we simplify the RCS calculation in the 3D space to \eqref{eq:rcs-scatterer-po-bis} by assuming that the RCSs of scatterer subpatch are identical for different azimuth angles. The assumption results in no significant influence on the results. More precise models of the RCS of rectangular PEC plates in 3D space can be found in \cite{balanis2012advanced}.} pointing from the center of the $(k,m)$-th subpatch to BS and UE, respectively.

Under the far-field assumption, we have $\vartheta_{k,m}^{{\rm{r}}} \approx \vartheta_{{\rm{r}}}$ and $\vartheta_{k,m}^{{\rm{t}}}\approx \vartheta_{{\rm{t}}}$, as shown in Fig. \ref{fig:structure-scatterer-ris}(a).
$\vartheta_{{\rm{r}}}$ and $\vartheta_{{\rm{t}}}$ represent the scattering and incident elevation angles pointing from the center of the PEC scatterer to BS and UE, respectively.
Thus, we obtain $\sigma_{k,m}^{\rm{subpatch}}\left(\vartheta_{k,m}^{{\rm{r}}}, \vartheta_{k,m}^{{\rm{t}}}\right)\approx \sigma_{\rm{subpatch}}\left(\vartheta_{{\rm{r}}}, \vartheta_{{\rm{t}}}\right)$.
Then, we can rewrite \eqref{eq:rcs-scatterer-summation-subpatch} as
\begin{equation}
\label{eq:rcs-scatterer-summation-subpatch-farfield}
\sigma_{\rm{s}} = \left|\sum_{k = 1}^K\sum_{m = 1}^M\sqrt{\sigma_{\rm{subpatch}}}e^{-j\alpha_{k,m}}\right|^2,
\end{equation}
where $\alpha_{k,m}$ is the phase shift of the path reflected by the $(k,m)$-th subpatch.
The location of the $(k,m)$-th subpatch is denoted as $\mathbf{p}_{k,m} = \left[x_{k,m},y_{k,m},z_{k,m}\right]^T$.
For simplicity, we assume that the center of the scatterer is located at the origin $\mathbf{p}_0 = [0,0,0]^T$.
Taking $\mathbf{p}_0$ as the reference point, we obtain $\alpha_{k,m} = \frac{ 2 \pi\left(\mathcal{d}_{k, m}^{\rm{r}}+\mathcal{d}_{k, m}^{\rm{t}}-d_{\rm{r}}-d_{\rm{t}}\right)}{\lambda}$.
Given that the scatterer is assumed to be in the far field of the UPAs of BS and UE, we have $\mathcal{d}_{k, m}^{\rm{r}} \approx d_{\rm{r}} - \mathbf{p}_{k,m} \mathbf{d}\left( \varphi_{\rm{r}}, \vartheta_{\rm{r}} \right)$ and $\mathcal{d}_{k, m}^{\rm{t}} \approx d_{\rm{t}} - \mathbf{p}_{k,m} \mathbf{d}\left( \varphi_{\rm{t}}, \vartheta_{\rm{t}} \right)$.
Therefore, the phase shift $\alpha_{k,m}$ is given by 
\begin{equation}
\label{eq:phase-shift-farfield}
\alpha_{k,m} = - \frac{2\pi}{\lambda} \left[ \mathbf{p}_{k,m}^T \mathbf{d}\left( \varphi_{\rm{r}}, \vartheta_{\rm{r}} \right) + \mathbf{p}_{k,m}^T \mathbf{d}\left( \varphi_{\rm{t}}, \vartheta_{\rm{t}} \right)\right].
\end{equation}

\subsection{RCS of RIS}

In the far fields, we take the similar approximation as \eqref{eq:rcs-scatterer-summation-subpatch-farfield} and rewrite \eqref{eq:nlos-pathgain-ris} as
\begin{equation}
\label{eq:nlos-pathgain-ris-farfield}
g^{\circ}_{n,l} = \frac{\lambda \sqrt{G G_{{\rm{R I S}}} F\left(\varphi_{{\rm{r}}}, \vartheta_{{\rm{r}}}\right) F\left(\varphi_{{\rm{t}}}, \vartheta_{{\rm{t}}}\right) d_{{\rm{x}}} d_{{\rm{y}}}}}{(4 \pi)^{\frac{3}{2}} d_{n,l}^{\rm{r}} d_{n,l}^{\rm{t}}} \times \sum_{k=1}^{K} \sum_{m=1}^{M} e^{-j\alpha_{k,m}} A e^{-j \xi_{k, m}},
\end{equation}
where $\alpha_{k,m}$ is in the same form as \eqref{eq:phase-shift-farfield}.
According to the definition of RCS in \eqref{eq:rcs-definition}, the following theorem gives the RCS of RISs that accounts for the wavelength and the scattering properties of RIS elements.
\begin{theorem}
\label{theorem:rcs-ris}
The RCS of RIS in the far-field region is as follows:
\begin{equation}
\label{eq:ris-rcs-whole}
\sigma_{\rm{RIS}} = \left|\sum_{k=1}^{K} \sum_{m=1}^{M} \sqrt{\sigma_{\rm{subRIS}}} e^{-j\left(\alpha_{k,m}+\xi_{k, m}\right)} \right|^2,
\end{equation}
where $\sigma_{\rm{subRIS}}$ is the RCS of each RIS element, which is given as
\begin{equation}
\label{eq:ris-rcs-sub}
\sigma_{\rm{subRIS}} =\frac{4\pi A^2S_{{\rm{sub}}}^2}{\lambda^2}F(\varphi_{\rm{r}}, \vartheta_{\rm{r}})F(\varphi_{\rm{t}}, \vartheta_{\rm{t}}).
\end{equation}
\end{theorem}

\emph{Proof}: See Appendix \ref{appendix:ris-prove}. \hfill $\blacksquare$

The comparison of \eqref{eq:rcs-scatterer-summation-subpatch-farfield} and \eqref{eq:ris-rcs-whole} reveals the similarities and differences between scatterers and RISs.
Both of their RCSs can be calculated by summing up the RCSs of scatterer subpatches or RIS elements according to the phase shifts caused by interelement path length differences.
However, the RCSs of RIS elements can be added constructively by controlling element phase shifts, whereas the RCSs of scatterer subpatches may be added destructively for most incident/scattering angles.
The RCSs of a single scatterer subpatch in \eqref{eq:rcs-scatterer-po-bis} and an RIS element in \eqref{eq:ris-rcs-sub} consider similar parameters and lead to the same expression as $\sigma = 4\pi S_{\rm{sub}}^2/\lambda^2$ when $\vartheta_{\rm{r}}=\vartheta_{\rm{t}}=0$ and $A=1$.
Moreover, the scattering directivity of RIS elements, i.e., the value of $\beta$ in \eqref{eq:pattern}, influences the RCS of RISs.
The radiation patterns of RIS elements can be designed to acquire the desired RCS at different incident/scattering angles.

Under the far-field assumption, we rewrite the path gain in \eqref{eq:nlos-pathgain-scatterer} and \eqref{eq:nlos-pathgain-ris-farfield} into the general model as
\vspace{-0.1cm}
\begin{equation}
\label{eq:nlos-pathgain-ris-scatterer}
g^{*\circ}_{n,l}=\frac{\lambda \sqrt{G \sigma} e^{-\frac{j 2 \pi (d_{n,l}^{\rm{r}}+d_{n,l}^{\rm{t}})}{\lambda}}}{(4 \pi)^{\frac{3}{2}} d_{n,l}^{\rm{r}} d_{n,l}^{\rm{t}}},
\end{equation}
where $\sigma$ represents the RCS of scatterers or RISs.
This general model verifies the effectiveness of treating RISs as a unique type of scatterer, by emphasizing the role of RCS.
Given that the optimal phase shifts provide a large RCS for RISs, a strong NLOS link can be established between the BS and UE, which overcomes the weakness of traditional NLOS paths and provides high-quality environmental information.
To conclude, Equations \eqref{eq:rcs-scatterer-summation-subpatch-farfield}, \eqref{eq:ris-rcs-whole}, and \eqref{eq:nlos-pathgain-ris-scatterer} uncover the similarities and differences between scatterers and RISs from the aspect of RCS.

\subsection{Properties and Comparisons of RCS}

RCS properties are discussed in this subsection, and fair comparisons are conducted between scatterers and RISs.
Frequency $f=27$ GHz, and wavelength $\lambda=11$ mm.
We set $K=M=40$ for the scatterer and RIS, and the size of each scatterer subpatch or RIS element is $d_{\rm{x}}=d_{\rm{y}}=0.4\lambda$.
Then, the total length of each side is $16\lambda$ (17.8 cm). 
The reflecting coefficient of RIS elements $A=1$.
The index of the radiation patterns of RIS elements is set to $\beta = 1$, which is the same as that in \cite{tang2021path}.
Notably, the unit of RCS is in $\text{m}^2$, and we have $10\ \text{log}\ x\ (\text{dBm}^2)$ with $x\ (\text{m}^2)$.

\begin{figure}
\vspace{-0.25cm}
\centering
\captionsetup{font=footnotesize}
  \hspace{-0.5cm}
  \begin{minipage}{0.49\linewidth}
    \centerline{\includegraphics[width=\textwidth]{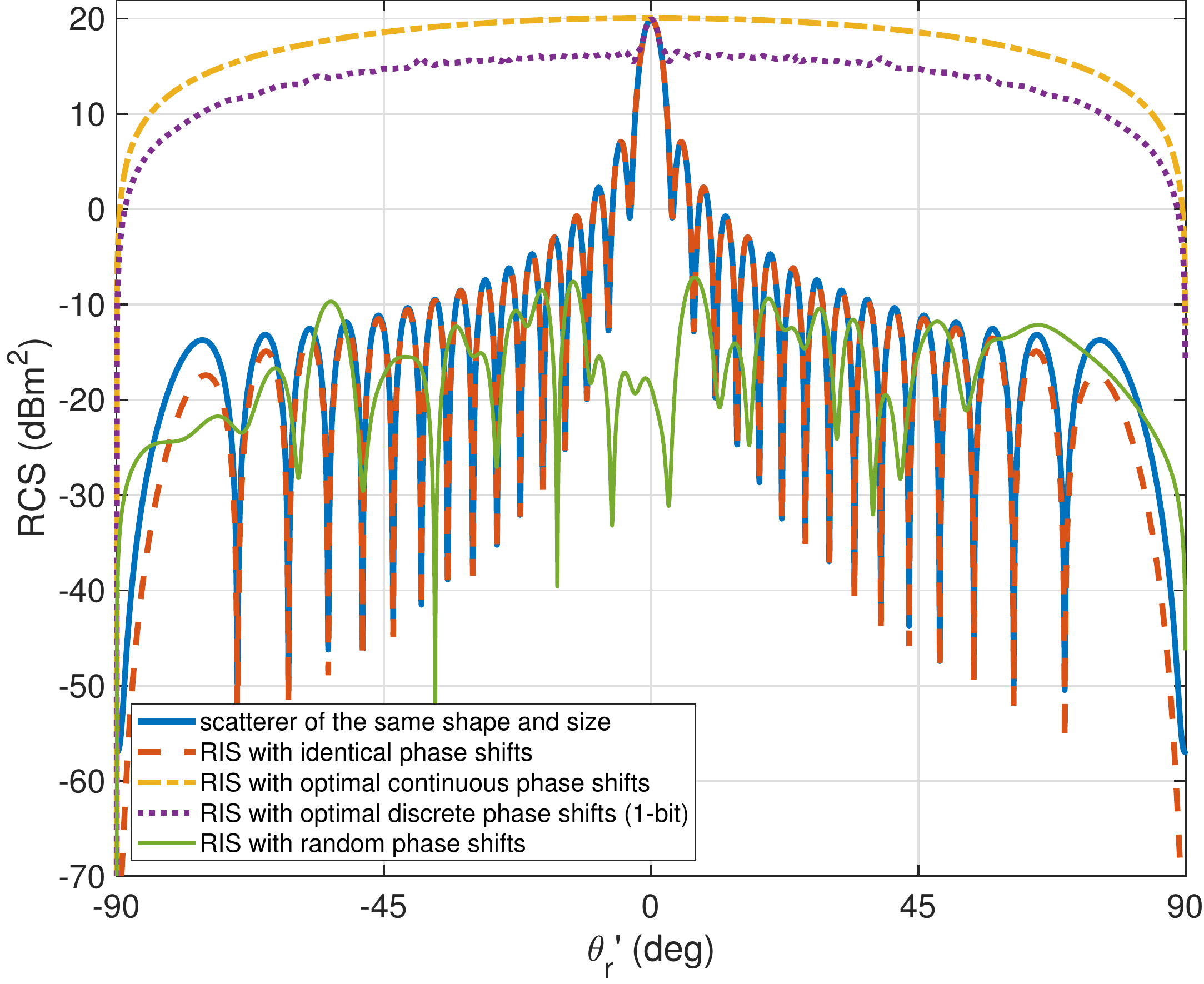}}
    \vspace{-0.1cm}
    \hspace{0.2cm}\centerline{\footnotesize{(a) $\vartheta_{\rm{t}}' = 0^\circ$}}
  \end{minipage}
  \hfill
  \begin{minipage}{0.49\linewidth}
    \centerline{\includegraphics[width=\textwidth]{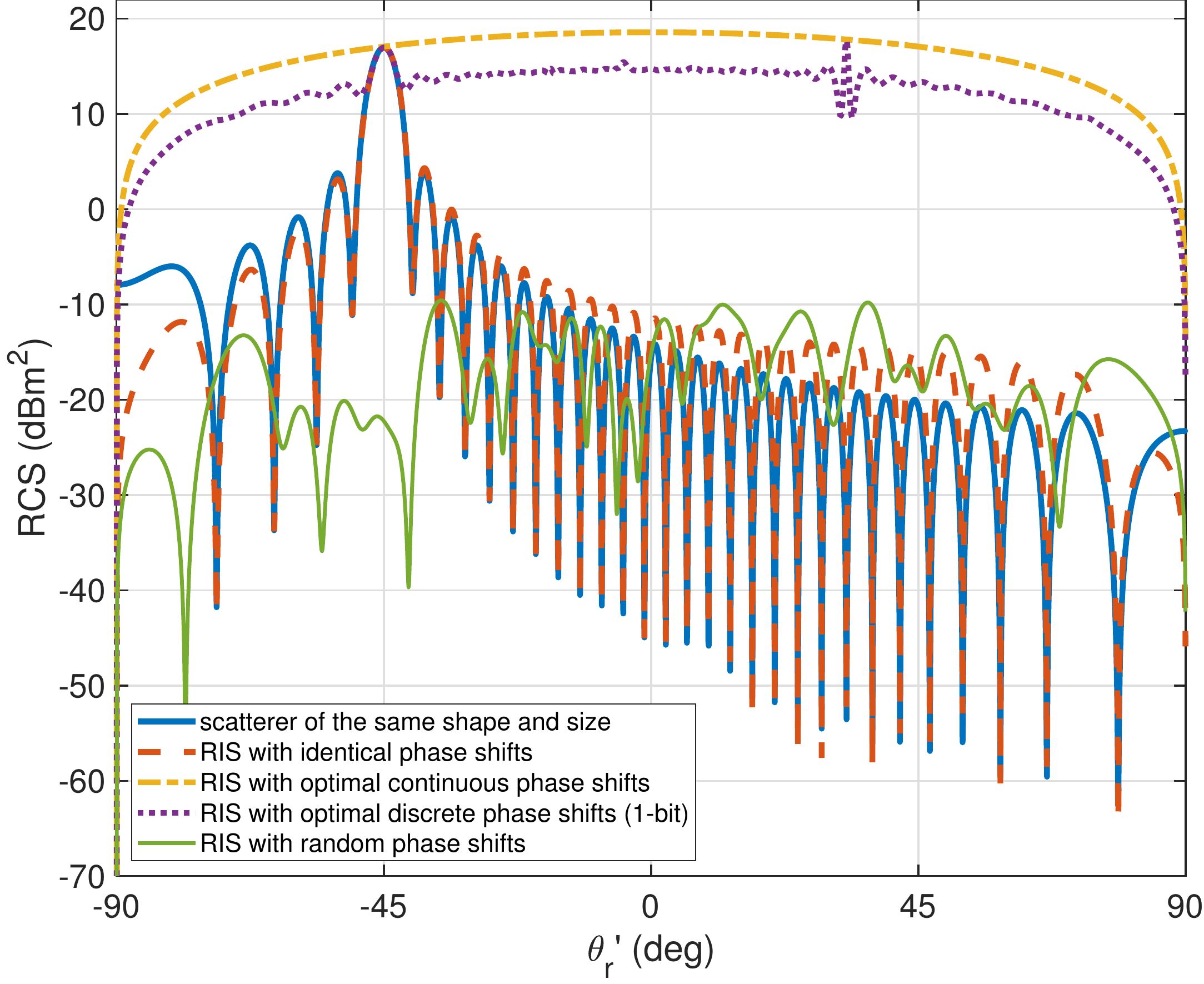}}
    \vspace{-0.1cm}
    \hspace{0.2cm}\centerline{\footnotesize{(b) $\vartheta_{\rm{t}}' = 45^\circ$}}
  \end{minipage}
  \vspace{0.2cm}
  \caption{RCS comparison of a scatterer and an RIS for various scattering angles.}
  \label{fig:simu-rcs-vs-angles}
  \vspace{-0.05cm}
\end{figure}

\subsubsection{RCS to Different Incident and Scattering Angles}

We set incident angles to be $\vartheta_{\rm{t}} = 0^\circ/45^\circ$.
To further reveal the scattering properties, we define $\vartheta_{\rm{r}}' = \vartheta_{\rm{r}}$, if $\varphi_{\rm{t}}-90^\circ \le \varphi_{\rm{r}} \le \varphi_{\rm{t}}+90^\circ$; otherwise, $\vartheta_{\rm{r}}' = - \vartheta_{\rm{r}}$.
In the simulation, $\vartheta_{\rm{r}}'$ varies from $-90^\circ$ to $90^\circ$, and Fig. \ref{fig:simu-rcs-vs-angles} displays the results.
Various configurations of the RIS are discussed: 
the RIS with identical phase shifts means that $\xi_{k,m}$ is identical for all elements, which is the same as a general metal plate;
optimal continuous phase shifts are given by $\xi_{k,m}^{\rm{con}} = - \alpha_{k,m}$, thus aligning the element phase shifts in-phase; 
the one-bit discrete phase shift means $\xi_{k,m}^{\rm{dis}} \in \{0, \pi\}$, and the optimal discrete phase shift is given by the nearest value to $\xi_{k,m}^{\rm{con}}$; 
random phase shifts are set stochastically.
The RIS with optimal continuous phase shifts can maintain a large RCS for most scattering angles because the RCSs of separate RIS elements are aligned in-phase, and we derive $\sigma_{\rm{RIS}}\!=\!(KM)^2\sigma_{\rm{subRIS}}$ from \eqref{eq:ris-rcs-whole}.
Even if only two discrete phase shifts are available, the RIS still achieves significant RCS enhancements compared with the scatterer.
Meanwhile, the results of optimal continuous phase shifts act as the upper bound of RCSs.
Moreover, the RCS of the scatterer with the same shape and size is normally much lower than the RISs discussed above, given that the phase shifts of scatterer subpatches are not aligned.
The RIS configured with identical phase shifts shares the same RCS fluctuations to the scattering angles with the scatterer, where the peak values are achieved at the specular angles, i.e., $\vartheta_{\rm{r}}'=-\vartheta_{\rm{t}}$, and decrease rapidly to less than 0 $\text{dBm}^2$ with $\vartheta_{\rm{r}}'$ deviating from $- \vartheta_{\rm{t}}$\footnote{Notably, the results may be incorrect for scatterers when the scattering angles are far from the specular angles, due to the neglect of edge effects in PO method \cite{Merrill1990radar,balanis2012advanced}. Similar insights for the RIS with identical phase shifts can also be found from the comparison of received signal power between the simulation and measurement results in \cite{tang2021path}.}.
Last, the random configuration of phase shifts obtains an even lower RCS than the scatterer.

\begin{figure}[t]
	\vspace{-0.25cm}
	\centering
	\captionsetup{font=footnotesize}
	\includegraphics[width=0.49\textwidth]{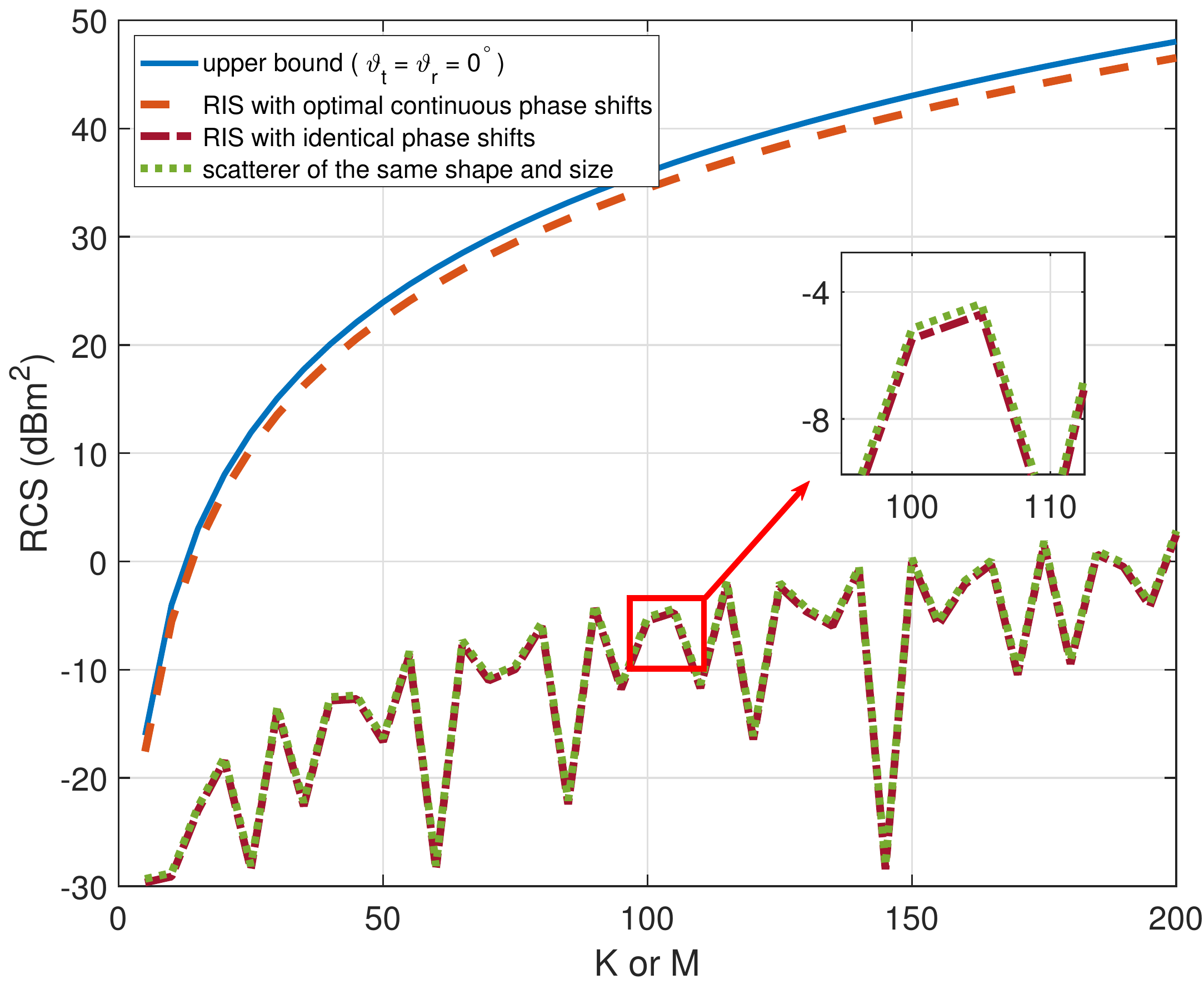}
	\vspace{0.2cm}
	\caption{RCS comparison of a scatterer and an RIS with different sizes ( $\vartheta_{\rm{t}}=0^\circ$, $\vartheta_{\rm{r}} = 45^\circ$ ).}
	\vspace{-0.2cm}
	\label{fig:simu-rcs-vs-size}
\end{figure}

\subsubsection{RCS to Different Sizes} 

We set $K=M$ to explore the effects of sizes on RCS.
Provided that the largest RCS for the scatterer and RIS is achieved when $\vartheta_{\rm{t}}=\vartheta_{\rm{r}}=0^\circ$ (easily derived from \eqref{eq:rcs-scatterer-po-bis} and \eqref{eq:ris-rcs-sub}), we take it as the upper bound of RCS in this scenario.
According to the results illustrated in Fig. \ref{fig:simu-rcs-vs-size}, the upper bound monotonously increases with the number of subpatches.
$\vartheta_{\rm{t}}=0^\circ$ and $\vartheta_{\rm{r}} = 45^\circ$ are taken as examples to derive the rest of the simulation results.
The same trend of monotonous increase is also discovered on the RCS of RIS configured with optimal continuous phase shifts, but the RCS is slightly reduced compared with the upper bound due to the deviation of $\vartheta_{\rm{r}}$ from $- \vartheta_{\rm{t}}$.
However, the RCSs of RIS with identical phase shifts and scatterer of the same shape and size are much smaller than the upper bound.
They share approximately the same fluctuations with the increase in size, although an increasing trend occurs. 
This phenomenon can be partly explained by the undulant shape of the $\text{Sa}(\cdot)$ function in \eqref{eq:rcs-scatterer-po-bis}.
Thus, a larger scatterer may result in a smaller RCS at non-specular angles.

\begin{remark}
The comparison of RISs and traditional scatterers shows that the RIS with identical phase shifts has similar scattering characteristics to a general PEC scatterer. 
However, the RIS has the advantage of maintaining a strong NLOS path due to its tunable element phase shifts, thus improving the signal-to-noise ratio (SNR) of NLOS paths and overcoming the drawback of traditional scatterers.
\end{remark}

In conclusion, the RIS can assist in joint localization and environment sensing by taking advantage of its improved NLOS paths \cite{shahmansoori2017position,mendrzik2018harnessing}. In the next section, we focus on this aspect of the study.

\section{Joint Localization and Environment Sensing}
\label{sec:wls}

\begin{figure}
	\vspace{-0.2cm}
	\centering
	\includegraphics[width=\textwidth]{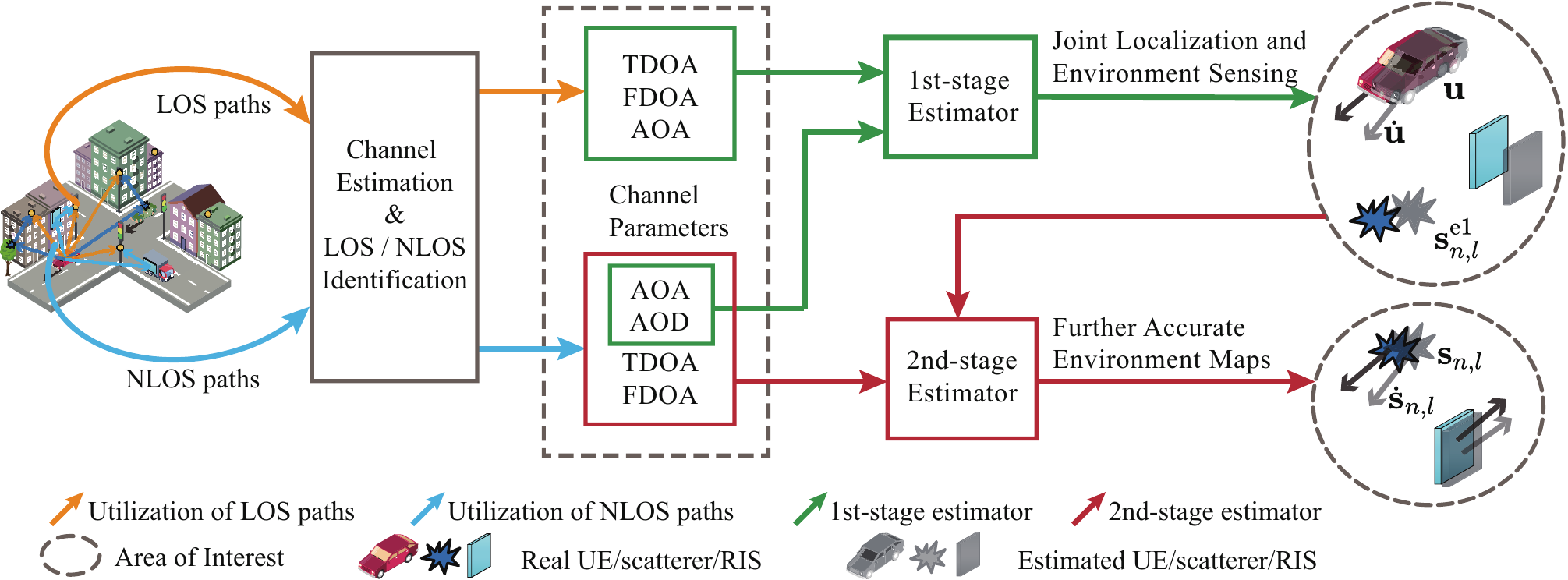}	
	\captionsetup{font=footnotesize}
	\vspace{-0.3cm}
	\caption{Illustration of the inputs and outputs of the proposed WLS estimators.}
	\label{fig:wls-flow}
\end{figure}

In this section, we propose the one-stage WLS estimator that jointly performs localization and environment sensing by exploiting NLOS components from scatterers and RISs.
First, LOS/NLOS paths are estimated and identified.
Second, the TDOA/FDOA/AOA measurements of LOS paths and AOA/AOD measurements of NLOS paths are used in a one-stage WLS estimator to estimate UE locations, velocities, and scatterer locations simultaneously. 
The second-stage estimator is employed to further enhance environment sensing accuracy by incorporating the outputs of the first-stage estimator and NLOS path measurements. 
The flow of the proposed estimators is illustrated in Fig. \ref{fig:wls-flow}, and they are executed on the same hardware platforms and spectrum resources utilized for communication.
RISs are treated as a unique type of scatterer, and they are not differentiated in the proposed estimators.
The localization process is the same for each UE in multi-user scenarios.

\subsection{CRLB of Channel Estimation}
\label{sec:crlb-channel-estimation}

We assume that the channel parameters given in Sec. \ref{sec:relation-parameter} have been estimated with advanced channel estimation techniques \cite{han2019efficient,guo2017millimeter,alkhateeb2014channel,zhu2017auxiliary} in this section.
Given that a CRLB gives a lower bound of the variance of any unbiased estimator \cite{kay1993fundamentals}, and the advanced channel estimation techniques can achieve the CRLB under low noise levels \cite{han2019efficient}, 
we treat the CRLB of channel estimation as the noise variance of estimated parameters 
$\eta_{n,l} = (\phi_{n,l}^{\rm{r}}, \theta_{n,l}^{\rm{r}}, \phi_{n,l}^{\rm{t}}, \theta_{n,l}^{\rm{t}}, \tau_{n,l}, \nu_{n,l})$, for $n=1, 2, \dots, N$ and $l=0, 1, \dots, L_n$. 
Then, the performance improvements on UE localization of scatterers and RISs are compared through simulation in Sec. \ref{sec:results-ris-aided}.

We stack the uplink multipath channels on all antennas, subcarriers, and symbols in \eqref{eq:channel-model-matrix} into a vector as \cite{han2019efficient}
\begin{equation}
\mathbf{h}_{n} =\sum_{l=0}^{L_{n}} g_{n,l}^\circ \underbrace{ \mathbf{a}_{\rm{r}}(\phi^{ {\rm{r}}\circ}_{n, l}, \theta^{ {\rm{r}}\circ}_{n, l}) \otimes \mathbf{a}_{\rm{t}}(\phi^{{\rm{t}}\circ}_{n, l}, \theta^{ {\rm{t}}\circ}_{n, l}) \otimes \mathbf{p}(\tau_{n,l}^\circ) \otimes \mathbf{q}(\nu_{n,l}^\circ)}_{\mathbf{w}(\phi^{ {\rm{r}}\circ}_{n, l}, \theta^{ {\rm{r}}\circ}_{n, l}, \phi^{ {\rm{t}}\circ}_{n, l}, \theta^{ {\rm{t}}\circ}_{n, l}, \tau^\circ_{n,l}, \nu_{n,l}^\circ)} ,
\end{equation}
where 
\begin{equation}
\mathbf{p}(\tau^\circ_{n,l})\!=\!\left[1, e^{-j2\pi \Delta f \tau^\circ_{n,l}}, \ldots, e^{-j2\pi \Delta f \tau^\circ_{n,l}(H-1)}\right]^{H}\!\!,\ 
\mathbf{q}(\nu_{n,l}^\circ) \!=\! \left[1, e^{-j2\pi T_{\rm{sym}} \nu_{n,l}^\circ}, \ldots, e^{-j2\pi T_{\rm{sym}} \nu_{n,l}^\circ(R-1)}\right]^{H}\!\!;
\end{equation}
$\Delta f$ is the subcarrier spacing, and $T_{\rm{sym}}$ is the symbol duration;
$H$ and $R$ are the numbers of subcarriers and symbols, respectively.
The Fisher information matrix (FIM) of $\eta_{n,l}^\circ$ is given by $\mathbf{F}(\eta_{n,l}^\circ)=2 \frac{\left|g_{n,l}^\circ\right|^{2}}{\sigma_{{\rm{z}}}^{2}} \Re(\mathbf{\Psi})$, \footnote{Notably, the RCS in path gain is also a function of AOA and AOD, illustrated by the discussions in Sec. \ref{sec:rcs}. Thus, $\partial g_{n,l}^\circ/\partial \theta_{n,l}^{{\rm{t}}\circ}$ and $\partial g_{n,l}^\circ/\partial \theta_{n,l}^{{\rm{r}}\circ}$ should also be calculated.
However, they are relatively small compared to the large numbers of antennas, subcarriers, and symbols, due to the large distance $d_1$ and $d_2$ in \eqref{eq:nlos-pathgain-ris-scatterer} under the far-field assumption.
Thus, the partial derivative of $g_{n,l}^\circ$ is omitted for simplicity, with no significant influence on the results.}
where $\mathbf{\Psi}_{i,j} = \frac{\partial \mathbf{w}^H}{\partial \eta_{n,l}^\circ(i)}\frac{\partial \mathbf{w}}{\partial \eta_{n,l}^\circ(j)}$, 
and $\sigma_{\rm{z}}^2$ is the variance of the additive Gaussian white noise, which considers the impacts of multi-bounce signals and refraction effects, including MUI in multi-user scenarios.
The CRLB of the $i$-th channel parameter in $\eta_{n,l}^\circ$ is denoted by $\text{CRLB}_{\eta_{n,l}^\circ(i)} = \mathbf{F}_{i,i}^{-1}(\eta_{n,l}^\circ)$.
Thus, the noise variance of the estimated parameter $\eta_{n,l}(i)$ is given as $\sigma_{\eta_{n,l}(i)}^2 = \text{CRLB}_{\eta_{n,l}^\circ(i)}$.
Provided that the delays and Doppler shifts are utilized in the form of TDOA- and FDOA-related parameters, the variances of $r_{n 1,l}$ and $\dot{r}_{n 1,l}$ can be written as 
$\sigma_{r_{n 1,l}}^2 = v^2(\sigma_{\tau_{n,l}}^2+\sigma_{\tau_{1,0}}^2)$ and $\sigma_{\dot{r}_{n 1,l}}^2 = \lambda^2(\sigma_{\nu_{n,l}}^2+\sigma_{\nu_{1,0}}^2)$, respectively.

Performing a measurement selection process to identify LOS/NLOS paths is necessary after channel estimation. 
This process can be achieved by comparing the received power of each path, and techniques such as the geometry-based algorithm in \cite{yang2021model} or machine learning-enabled method in \cite{huang2020machine} can be employed. 
However, as this topic is not the main focus of this study, we assume that LOS and NLOS path measurements have already been separated.

\subsection{One-stage Estimator for UE and Scatterers}
\label{sec:wls-1-stage}

In this subsection, we present a closed-form one-stage WLS estimator to jointly estimate UE locations, velocities, and scatterer locations. 
Unlike the traditional methods in \cite{yang2021model}, which rely solely on LOS path measurements, our proposed estimator leverages LOS and NLOS paths, where 
NLOS path measurements not only realize scatterer localization, but also assist in UE localization.

We assume that $N$ BSs simultaneously connect to UE, and $L_{n}$ scatterers or RISs exist between the $n$-th BS and UE, where $n\!=\!1,2,\ldots,N$.
Provided that the TDOA/FDOAs of LOS paths take the TOA/FOA of the first BS as references, we have $N\!-\!1$ TDOAs and $N\!-\!1$ FDOAs.
The following parameters also exist: $2N$ AOAs of LOS paths, $2N_{\rm{s}}$ AOAs of NLOS paths, and $2N_{\rm{s}}$ AODs of NLOS paths, where $N_{\rm{s}}\!=\!\sum_{n=1}^N L_n$.
Then, the noise-free parameters are denoted as $\mathbf{m}^{\circ} = \left[{\mathbf{m}^{\circ}_1}^T,{\mathbf{m}^{\circ}_2}^T\right]^{T}$, 
where $\mathbf{m}^{\circ}_1\!=\!\left[r_{21,0}^{\circ}, \dot{r}_{21,0}^{\circ}, \ldots, r_{N 1,0}^{\circ}, \dot{r}_{N 1,0}^{\circ}, \phi_{1,0}^{{\rm{r}}\circ}, \theta_{1,0}^{{\rm{r}}\circ}, \ldots, \phi_{N,0}^{{\rm{r}}\circ}, \theta_{N,0}^{{\rm{r}}\circ}\right]^{\!T}$ 
and $\mathbf{m}^\circ_2 = \left[{\mathbf{m}^{\circ T}_{2,1}}, {\mathbf{m}^{\circ T}_{2,2}}, \ldots, {\mathbf{m}^{\circ T}_{2,N}}\right]^T$
represent the LOS and NLOS path parameters, respectively.
Here, $\mathbf{m}^\circ_{2,n}=\left[\phi_{n,1}^{{\rm{r}}\circ},\theta_{n,1}^{{\rm{r}}\circ},\phi_{n,1}^{{\rm{t}}\circ},\theta_{n,1}^{{\rm{t}}\circ},\ldots,\phi_{n,L_{n}}^{{\rm{r}}\circ},\theta_{n,L_{n}}^{{\rm{r}}\circ},\phi_{n,L_{n}}^{{\rm{t}}\circ},\theta_{n,L_{n}}^{{\rm{t}}\circ}\right]^T$. 
In summary, $4N-2+4N_{\rm{s}}$ parameters are available for the estimator. 
Next, we first employ the noise-free parameters to establish pseudo linear equations with respect to unknown $\mathbf{u}^\circ$, $\dot{\mathbf{u}}^\circ$, and $\mathbf{s}^\circ_{n,l}$.
Then, we apply the WLS estimator to realize joint localization and environment sensing with the existence of measurement noise.

The parameters in $\mathbf{m}^\circ$ can be utilized to establish $4N-2+4N_{\rm{s}}$ pseudo linear equations, which are given in the matrix form as
\begin{equation}\label{eq:wls-1-collected-matrix}
\mathbf{h}^\circ=\mathbf{G}^\circ\mathbf{x}^\circ,
\vspace{-0.3cm}
\end{equation}
where 
\begin{equation}
\begin{array}{c@{}l}
\mathbf{h}^\circ = 
\left[
\begin{array}{c}
\mathbf{h}_{\rm{u}}^\circ  \\
\hdashline[2pt/2pt] 
\mathbf{h}_{\rm{s}}^\circ \\
\end{array}
\right]
\begin{array}{l}
\left.\rule{-3 mm}{4 mm}\right\} {\scriptstyle 4N-2 }\\
\left.\rule{-3 mm}{4 mm}\right\} {\scriptstyle 4N_{\rm{s}}}
\end{array}\\[-13pt]
\begin{array}{ccc}
\;
\;
\end{array}
\end{array}, \quad
\begin{array}{c@{}l}
\mathbf{G}^\circ = 
\left[
\begin{array}{c;{2pt/2pt}c}
\mathbf{G}_{\rm{u}}^\circ & \mathbf{O} \\
\hdashline[2pt/2pt] 
\mathbf{G}_{\rm{s1}}^\circ & \mathbf{G}_{\rm{s2}}^\circ\\
\end{array}
\right]
\begin{array}{l}
\left.\rule{-3 mm}{4 mm}\right\} {\scriptstyle 4N-2 }\\
\left.\rule{-3 mm}{4 mm}\right\} {\scriptstyle 4N_{\rm{s}}}
\end{array}\\[-13pt]
\begin{array}{ccc}
\;
\!\!\!\!\!\underbrace{\rule{10 mm}{-1 mm}}_6
\underbrace{\rule{10 mm}{-1 mm}}_{3N_{\rm{s}}}
\;
\end{array}
\end{array};
\end{equation}
$\mathbf{x}^\circ=[\mathbf{u}^\circ;\dot{\mathbf{u}}^\circ;\mathbf{s}^\circ]$ is the unknown 
$(6+3N_{\rm{s}})$-dimensional state vector of UE and scatterers;
$\mathbf{s}^\circ=\left[{\mathbf{s}_1^\circ}; \ldots; {\mathbf{s}_{N}^\circ}\right]$ and 
$\mathbf{s}_n^\circ=\left[{\mathbf{s}_{n,1}^\circ}; \ldots; {\mathbf{s}_{n,L_{n}}^\circ}\right]$ represent the scatterer locations.
In \eqref{eq:wls-1-collected-matrix}, the first $4N-2$ equations, which are specified by $\mathbf{h}_{\rm{u}}^\circ$ and $\mathbf{G}_{\rm{u}}^\circ$, are established by the LOS path parameters in $\mathbf{m}^{\circ}_1$.
In the case of TDOA parameter $r_{n1,0}^\circ$, we rewrite \eqref{eq:tdoa} as $r_{n1,0}^\circ+r_{1,0}^\circ=r_{n,0}^\circ$, with $n=2,3,\ldots,N$. 
Squaring both sides, substituting $r_{n,0}^\circ$ by \eqref{eq:tdoa-goe}, and multiplying both sides by $\mathbf{d}^T\left(\phi_{1,0}^{\rm{r}\circ}, \theta_{1,0}^{\rm{r}\circ}\right) \mathbf{d}\left(\phi_{1,0}^{\rm{r}\circ}, \theta_{1,0}^{\rm{r}\circ}\right)$, we obtain $N-1$ pseudo linear equations with respect to $\mathbf{u}^\circ$, which are given as \eqref{eq:tdoa-wls-eq} in Appendix \ref{appendix:wls-equations}.
Taking the time derivative of these TDOA-related equations yields another $N-1$ FDOA-related equations with respect to $\mathbf{u}^\circ$ and $\dot{\mathbf{u}}^\circ$, given as \eqref{eq:fdoa-wls-eq}.
Moreover, the vector pointing from the $n$-th BS to the UE is given by $\mathbf{u}^\circ - \mathbf{b}_n$, whose unit-norm vector is denoted by $\mathbf{d}\left(\phi_{n,0}^{\rm{r}\circ},\theta_{n,0}^{\rm{r}\circ}\right)$ and is orthogonal with $\mathbf{c}\left(\phi_{n,0}^{\rm{r}\circ},\theta_{n,0}^{\rm{r}\circ}\right)$ and $\mathbf{v}\left(\phi_{n,0}^{\rm{r}\circ},\theta_{n,0}^{\rm{r}\circ}\right)$. Thus, $2N$ AOA-related equations with respect to $\mathbf{u}^\circ$ can be established, given as \eqref{eq:aoa-wls-eq-los}.
The last $4N_{\rm{s}}$ equations specified by $\mathbf{h}_{\rm{s}}^\circ$, $\mathbf{G}_{\rm{s}1}^\circ$, and $\mathbf{G}_{\rm{s}2}^\circ$ are derived with the NLOS path parameters in $\mathbf{m}^{\circ}_2$.
Considering the similar definition of AOA and AOD in \eqref{eq:los-aoa}, \eqref{eq:nlos-aoa}, and \eqref{eq:nlos-aod}, the pseudo linear equations of AOA/AODs of NLOS paths are similar to those of the AOAs of LOS paths.
Specifically, $2N_{\rm{s}}$ AOAs yield $2N_{\rm{s}}$ equations with respect to $\mathbf{s}^\circ$, and $2N_{\rm{s}}$ AODs correspond to $2N_{\rm{s}}$ equations with respect to $\mathbf{s}^\circ$ and $\mathbf{u}^\circ$, given as \eqref{eq:aoa-aod-wls-eq-nlos}.

The proposed estimator is distinct from previous studies \cite{yang2021model,yu2021efficient,wang2015asymptotically,ho2004accurate,amiri2017asymptotically,amiri2017efficient} by incorporating NLOS path measurements. 
These measurements not only provide information for environment sensing, allowing scatterer location estimation, but also contribute to UE localization. 
Unlike the WLS estimator in \cite{yang2021model}, the proposed estimator determines UE locations, velocities, and scatterer locations simultaneously, without relying on prior knowledge about UE locations and velocities. 
The presence of non-zero matrix $\mathbf{G}_{\rm{s1}}^\circ$ in \eqref{eq:wls-1-collected-matrix} indicates that NLOS path parameters also provide information about UE locations. 
As a result, the proposed estimator with NLOS paths leads to improved UE localization accuracy compared with the estimators that only use LOS paths. 
Environment sensing accuracy is also enhanced, as discussed in the following subsection.

Next, we present the closed-form one-stage WLS estimator to perform joint localization and environment sensing in the presence of measurement noise.
The hybrid measurements with the additive noise are modeled as $\mathbf{m}=\mathbf{m}^{\circ}+\Delta \mathbf{m}$, where $\Delta \mathbf{m}$ is a Gaussian noise vector with zero mean and covariance matrix $\mathbf{Q}$.
Replacing the parameters in \eqref{eq:wls-1-collected-matrix} with noisy ones, we obtain the error vector 
\vspace{-0.1cm}
\begin{equation}\label{eq:wls-1-error-vector}
\mathbf{e}={\mathbf{h}}-{\mathbf{G}} \mathbf{x}^{\circ}, 
\vspace{-0.3cm}
\end{equation}
where ${\mathbf{h}}$ and ${\mathbf{G}}$ are the noisy forms of $\mathbf{h}^\circ$ and $\mathbf{G}^\circ$.
The solution of $\mathbf{x}^\circ$ is obtained by \cite{kay1993fundamentals}
\vspace{-0.1cm}
\begin{equation}\label{eq:wls-1-answer}
\mathbf{x}=\left({\mathbf{G}}^{T} \mathbf{W} {\mathbf{G}}\right)^{-1} {\mathbf{G}}^{T} \mathbf{W} {\mathbf{h}},
\vspace{-0.2cm}
\end{equation}
where $\mathbf{W}=(\mathbb{E}\{\mathbf{e}\mathbf{e}^T\})^{-1}$ is a positive definite weighting matrix.
Owing to the nonlinearity of $\mathbf{e}$, directly obtaining $\mathbf{W}$ is normally difficult.
Thus, we ignore the second- and high-order noise terms in $\mathbf{e}$ under low noise levels and carry out linear approximation by taking the Taylor expansion of \eqref{eq:wls-1-error-vector}.
Then, we obtain
\vspace{-0.1cm}
\begin{equation}\label{eq:wls-1-error-vector-approx}
\mathbf{e} \approx \mathbf{B} \Delta \mathbf{m},
\vspace{-0.3cm}
\end{equation}
where matrix $\mathbf{B}$ is the approximating coefficients of noise vector $\Delta \mathbf{m}$, and its detailed form is given as \eqref{eq:matrix-b-wls-1} in Appendix \ref{appendix:wls-equations}.
Therefore, vector $\mathbf{e}$ is approximated by a zero-mean Gaussian vector with covariance matrix $\mathbf{B}\mathbf{Q}\mathbf{B}^T$.
Thus, the weighting matrix $\mathbf{W}$ that minimizes the variance of estimate ${\mathbf{x}}$ is given by
\vspace{-0.1cm}
\begin{equation}\label{eq:wls-1-weighting-matrix}
\mathbf{W} =\left(\mathbf{B}\mathbf{Q}\mathbf{B}^T\right)^{-1}.
\vspace{-0.2cm}
\end{equation}
Considering that matrix $\mathbf{B}$ is unknown at the beginning, we initialize $\mathbf{W}^{(0)}=\mathbf{Q}^{-1}$ to provide primary estimates.
Then, we can obtain a rough estimate of $\mathbf{B}$ and renew the weighting matrix by \eqref{eq:wls-1-weighting-matrix}.
One or two iterative operations, such as this one, can attain the estimates with high accuracy.

Finally, we evaluate the performance of the proposed estimator by deriving the CRLB of $\mathbf{x}^\circ$, which is given in Appendix \ref{sec:crlb-wls-1}.
Then, we prove that the estimator is asymptotically unbiased, and the covariance matrix $\text{cov}(\mathbf{x}) \approx \left(\left(\mathbf{B}^{-1} \mathbf{G}\right)^{T} \mathbf{Q}^{-1} \mathbf{B}^{-1} \mathbf{G}\right)^{-1}$ approaches the CRLB with low measurement noise in Appendix \ref{sec:cov-wls-1}.
RISs enhance localization and environment sensing performances by tuning the phase shifts of their elements.
As discussed in Sec. \ref{sec:crlb-channel-estimation}, $\sigma_{\eta_{n,l}(i)}^2$ is proportional to $\frac{\sigma_{{\rm{z}}}^{2}}{\left|g_{n,l}^\circ\right|^{2}}$.
The variance of the measurements on the $(n,l)$-th path (i.e., the elements in matrix $\mathbf{Q}$) is proportional to $\frac{\sigma_{{\rm{z}}}^{2}}{\left|g_{n,l}^\circ\right|^{2}}$.
According to Equation \eqref{eq:crlb-final}, we have $\mathbf{C R L B}\left(\mathbf{x}^{\circ}\right) \propto \frac{\sigma_{{\rm{z}}}^{2}}{\left|g_{n,l}^\circ\right|^{2}}$, where $\propto$ means the positive correlation.
Therefore, the improved NLOS path gain by the RIS degrades the CRLB of $\mathbf{x}^\circ$.

\begin{remark}
\label{remark:new-response1}
RIS element phase shifts should be properly determined to enhance NLOS paths. 
This task can be achieved by defining the optimal phase shifts that achieve the largest RCS as a function of the locations of UE and RISs, given as $\mathcal{F}(\mathbf{u}^\circ, \mathbf{s}^\circ)$ \cite{tang2021path}.
A rough estimate of the locations of UE and RISs can be obtained by initializing the phase shifts randomly. 
Then, the optimal phase shifts for the present estimate can be determined by evaluating $\mathcal{F}(\mathbf{u}^\circ, \mathbf{s}^\circ)$ at the initial estimate of the locations. 
In addition, some RISs can be deployed on mobile vehicles \cite{huang2021transforming}, which can adjust the phase shifts based on the feedback from BSs, allowing for more accurate location estimates. 
Repeating this process iteratively, the RIS phase shifts can be gradually tuned to approach the optimal values, and the location estimates can be improved over time \cite{wang2022location}.
\end{remark}
\vspace{-0.5cm}

\begin{remark}
\label{remark:wls-1-special-case}
The proposed estimator can perform joint localization and environment sensing in conditions where only NLOS paths are available, but two conditions must be met. 
First, the number of scatterers (NLOS paths) must be at least three, meaning that the number of measurements ($4N_{\rm{s}}$) must be equal to or greater than the number of unknown parameters ($3+3N_{\rm{s}}$). 
Second, NLOS paths must come from two or more BSs, so that distance information can be obtained from known BS locations. 
If the NLOS paths are from a single BS, the estimator may not function as expected, because matrix $\mathbf{G}^\circ$ in \eqref{eq:wls-1-collected-matrix} will neither have a full column rank nor provide the necessary information for localization. 
That is, UE and scatterers can be located with only NLOS path measurements if three or more scatterers are linked to at least two BSs.
\end{remark}

\subsection{Second-stage Estimator for Scatterers}
\label{sec:wls-2-stage}

Inspired by the work of \cite{yang2021model}, the TDOA/FDOA parameters of NLOS paths can also be used to estimate scatterer locations.
However, the equations of TDOA/FDOAs, which will be established later, are non-linear functions of unknown UE locations and velocities.
Consequently, utilizing TDOA/FDOAs in the proposed one-stage estimator is difficult.
To further improve environment sensing accuracy, a second-stage estimator is proposed in this subsection by using the estimated UE locations and velocities.
Moreover, the rough estimates of scatterer locations in the first-stage estimator can help obtain scatterer locations with high accuracy.

Then, we show how the second-stage estimator works.
Given that the estimator can be independently employed for each scatterer or RIS, we only take the $(n,l)$-th scatterer as an example in this subsection.
First, the noise-free parameters of the $(n,l)$-th NLOS path that are utilized in the second-stage estimator are denoted by $\mathbf{m}_{n, l}^{\circ}\!=\!\left[r_{n 1, l}^{ \circ}, \dot{r}_{n 1, l}^{ \circ}, \phi_{n, l}^{{\rm{r}} \circ}, \theta_{n, l}^{{\rm{r}} \circ}\right.$, $\left.\phi_{n, l}^{{\rm{t}} \circ}, \theta_{n, l}^{{\rm{t}} \circ}\right]^{T}\!\!$, 
whose noisy counterparts are $\mathbf{m}_{n, l}=\mathbf{m}_{n, l}^{\circ}+\Delta \mathbf{m}_{n, l}$, where $\Delta \mathbf{m}_{n, l}^{{\rm{s}} }$ is zero mean Gaussian noise with covariance matrix $\mathbf{Q}_{n,l}^{{\rm{s}}}$.
Compared with the estimator in Sec. \ref{sec:wls-1-stage}, we further introduce TDOA/FDOA parameters into the second-stage estimator, which provides new information about scatterer locations.
Thus, improved localization accuracy is anticipated.
Second, we denote the rough estimate of the $(n,l)$-th scatterer in the first stage as $\mathbf{s}_{n,l}^{{\rm{e}} 1}= \left[x_{n,l}^{{\rm{s,e}}1}, y_{n,l}^{{\rm{s,e}}1}, z_{n,l}^{{\rm{s,e}}1}\right]^T$.
The estimation error of $\mathbf{s}_{n,l}^{{\rm{e}} 1}$ is given by $\Delta\mathbf{s}_{n,l}^{{\rm{e}} 1}=\mathbf{s}_{n,l}^{{\rm{e}} 1}-\mathbf{s}_{n,l}^{\circ}$, whose covariance matrix can be approximated by the CRLB under low noise levels and is given by $\mathbf{Q}_{n,l}^{{\rm{e}}1}$.
The rough estimate acts as a constraint of the scatterer location in the second-stage estimator and further improves location accuracy.
$\mathbf{m}_{n, l}$ and $\mathbf{s}_{n,l}^{{\rm{e}}1}$ are considered uncorrelated; 
thus, the covariance matrix of the noisy terms in the second-stage estimator is given by $\mathbf{Q}_{n,l} = \text{blkdiag}(\mathbf{Q}_{n,l}^{\rm{s}},\mathbf{Q}_{n,l}^{\rm{e}1})$.

Next, we establish pseudo linear equations with respect to $\mathbf{s}_{n,l}^\circ$ and $\dot{\mathbf{s}}_{n,l}^\circ$ by using the estimated UE location and velocity in Sec. \ref{sec:wls-1-stage}.
The parameters in $\mathbf{m}_{n, l}^{\circ}$ and the constraint $\mathbf{s}_{n,l}^{{\rm{e}} 1}$ help establish nine pseudo linear equations, given as
\begin{equation}\label{eq:wls-2-collected-matrix}
\mathbf{h}_{n, l}^{\rm{s}\circ}=\mathbf{G}_{n, l}^{\rm{s}\circ} \mathbf{x}_{n, l}^{\circ},
\vspace{-0.3cm}
\end{equation}
where 
\begin{equation}
\begin{array}{c@{}l}
\mathbf{h}_{n, l}^{\rm{s}\circ} = 
\left[
\begin{array}{c}
\mathbf{h}_{n, l}^{\rm{s1}\circ}  \\
\hdashline[2pt/2pt] 
\mathbf{h}_{n, l}^{\rm{s2}\circ} \\
\hdashline[2pt/2pt] 
\mathbf{s}_{n,l}^{{\rm{e}} 1}\\
\end{array}
\right]
\begin{array}{l}
\left.\rule{-3 mm}{4 mm}\right\} {\scriptstyle 2 }\\
\left.\rule{-3 mm}{4 mm}\right\} {\scriptstyle 4}\\
\left.\rule{-3 mm}{4 mm}\right\} {\scriptstyle 3}
\end{array}\\[-6pt]
\begin{array}{ccc}
\;
\;
\;
\end{array}
\end{array}, \quad
\begin{array}{c@{}l}
\mathbf{G}_{n, l}^{\rm{s}\circ} = 
\left[
\begin{array}{c;{2pt/2pt}c}
\mathbf{G}_{n, l}^{\rm{s11}\circ} & \mathbf{G}_{n, l}^{\rm{s12}\circ} \\
\hdashline[2pt/2pt] 
\mathbf{G}_{n, l}^{\rm{s2}\circ} & \mathbf{O}\\
\hdashline[2pt/2pt] 
\mathbf{I}_{3\times 3} & \mathbf{O}\\
\end{array}
\right]
\begin{array}{l}
\left.\rule{-3 mm}{4 mm}\right\} {\scriptstyle 2}\\
\left.\rule{-3 mm}{4 mm}\right\} {\scriptstyle 4}\\
\left.\rule{-3 mm}{4 mm}\right\} {\scriptstyle 3}
\end{array}\\[-13pt]
\begin{array}{ccc}
\;
\ \ \ \ \underbrace{\rule{10 mm}{-1 mm}}_3
\ \ \underbrace{\rule{10 mm}{-1 mm}}_{3}
\;
\end{array}
\end{array};
\end{equation}
$\mathbf{x}_{n,l}^{\circ}=\left[\mathbf{s}_{n,l}^{\circ},\dot{\mathbf{s}}_{n,l}^{\circ}\right]^T$ is the unknown location and velocity of the $(n,l)$-th scatterer.
The first two equations in \eqref{eq:wls-2-collected-matrix}, which are specified by $\mathbf{h}_{n, l}^{\rm{s1}\circ}$, $\mathbf{G}_{n, l}^{\rm{s11}\circ}$, and $\mathbf{G}_{n, l}^{\rm{s12}\circ}$, are established from the TDOA parameter $r_{n 1, l}^{ \circ}$ and FDOA parameter $\dot{r}_{n 1, l}^{ \circ}$.
Similar to the first-stage estimator, the first equation is derived from \eqref{eq:tdoa}, i.e., $r_{n1,l}^\circ + r_{1,0}^\circ - d_{n,l}^{{\rm{r}}\circ} = d_{n,l}^{{\rm{t}}\circ}$, where $d_{n,l}^{{\rm{r}}\circ}=\left\|\mathbf{s}_{n, l}^{\circ}-\mathbf{b}_{n}\right\|$, and 
$d_{n,l}^{{\rm{t}}\circ}=\left\|\mathbf{u}^{\circ}-\mathbf{s}_{n, l}^{\circ}\right\|$. 
$r_{1,0}^\circ$ is known because the UE location has been estimated.
Taking the time derivative of the first equation, we derive the second equation with respect to $\mathbf{s}^\circ_{n,l}$ and $\dot{\mathbf{s}}^\circ_{n,l}$.
The detailed forms of the first two equations are given as \eqref{eq:detailed-wls-2-eq1} and \eqref{eq:detailed-wls-2-eq2} in Appendix \ref{appendix:wls-equations}.
The following four equations (specified by $\mathbf{h}_{n, l}^{\rm{s2}\circ}$ and $\mathbf{G}_{n, l}^{\rm{s2}\circ}$) are related to AOA/AOD parameters, which are in the same forms as those in the first-stage estimator.
The last three equations consider the constraint $\mathbf{s}_{n,l}^{{\rm{e}} 1}$ and integrate the environmental information obtained in the first stage into the second-stage estimator.

Provided that one FDOA measurement is incapable of estimating 3D scatterer velocities, we take the same assumption as \cite{yang2021model} that the scatterers are vehicles moving along the same road as UE, thus rendering them in parallel velocity directions.
Defining the unit direction vector $\mathbf{n}_{\rm{v}}^\circ=\dot{\mathbf{u}}^\circ/\|\dot{\mathbf{u}}^\circ\|$, we have $\dot{\mathbf{s}}^\circ_{n,l}=\dot{s}^\circ_{n,l}\mathbf{n}_{\rm{v}}^\circ$, where $\dot{s}^\circ_{n,l}$ represents the velocity magnitude. 
Taking $\tilde{\mathbf{x}}_{n,l}^\circ=[\mathbf{s}_{n,l}^{\circ T},\dot{s}^\circ_{n,l}]^T$ as the new unknown state vector of the scatterer, we obtain
\vspace{-0.1cm}
\begin{equation}
\mathbf{h}_{n, l}^{\rm{s}\circ}=\mathbf{G}_{n, l}^{\rm{s}\circ} \mathbf{K} \tilde{\mathbf{x}}_{n, l}^{ \circ},
\vspace{-0.1cm}
\end{equation}
where $\mathbf{K} = \text{blkdiag}(\mathbf{I}_{3 \times 3}, \mathbf{n}_{{\rm{v}}}^\circ)$ is a $6 \times 4$ transformation matrix.
The solution of $\tilde{\mathbf{x}}_{n,l}^\circ$ is given by 
\begin{equation}
\tilde{\mathbf{x}}_{n, l}^{}=\left(\left({\mathbf{G}}_{n, l}^{\rm{s}} \mathbf{{K}}\right)^{T} \mathbf{W}_{n, l} {\mathbf{G}}_{n, l}^{\rm{s}} {\mathbf{K}}\right)^{-1}\left({\mathbf{G}}_{n, l}^{\rm{s}} \mathbf{{K}}\right)^{T} \mathbf{W}_{n, l} {\mathbf{h}}_{n, l}^{\rm{s}},
\end{equation}
where ${\mathbf{G}}_{n, l}^{\rm{s}}$ and ${\mathbf{h}}_{n, l}^{\rm{s}}$ are the noisy counterparts of ${\mathbf{G}}_{n, l}^{\rm{s}\circ}$ and ${\mathbf{h}}_{n, l}^{\rm{s}\circ}$, respectively, and weighting matrix $\mathbf{W}_{n, l}^{}=\left(\mathbf{B}_{n, l}^{} \mathbf{Q}_{n, l}^{} \mathbf{B}_{n, l}^T\right)^{-1}$. 
Matrix $\mathbf{B}_{n, l}$ is derived by executing linear approximation analogous to that in \eqref{eq:wls-1-error-vector-approx} and is omitted here due to space limitation.
The estimator is initialized by $\mathbf{W}_{n, l}^{{\rm{s}}(0)}=\left(\mathbf{Q}_{n, l}^{{\rm{s}}}\right)^{-1}$
and proven efficient under low noise levels, with covariance matrix $\text{cov}\left(\tilde{\mathbf{x}}_{n, l}\right) \!\approx\!\left(\left(\mathbf{B}_{n, l}^{-1} \mathbf{G}_{n, l}^{\rm{s}} \mathbf{K}\right)^{\!T} \!\mathbf{Q}_{n, l}^{-1} \mathbf{B}_{n, l}^{-1} \mathbf{G}_{n, l}^{\rm{s}} \mathbf{K}\right)^{-1}$.
Proofs are similar to that of the first-stage estimator.

\begin{remark}
The second-stage estimator verifies the benefit of the joint utilization of LOS and NLOS path measurements in the first stage to environment sensing.
With the rough estimate $\mathbf{s}_{n,l}^{{\rm{e}} 1}$ and covariance matrix $\mathbf{Q}_{n,l}^{{\rm{e}}1}$, the information about surrounding environments derived in the first stage is conveyed to the second-stage estimator. 
Incorporating $\mathbf{s}_{n,l}^{{\rm{e}} 1}$, $\mathbf{Q}_{n,l}^{{\rm{e}}1}$, and NLOS path parameters, scatterer location accuracy can be improved.
\end{remark}

\clearpage
\section{Numerical Results}
\label{sec:results}
\subsection{Performance of the Proposed WLS Estimator}

In this subsection, we first evaluate the performance of our proposed WLS estimators with given measurements and noise.
Given that BSs are densely deployed in mmWave systems, we set $N=6$ for the simulations.
Table \ref{tab:bs-location} lists the BS locations.
The environment comprises 18 scatterers. 
Table \ref{tab:scatterer-location} enumerates the scatterer locations and corresponding BSs, whereas different scatterer locations are discussed in Sec. \ref{sec:results-ris-aided}.
We denote the covariance matrix of noise $\Delta\mathbf{m}$ as
\begin{equation}
\mathbf{Q}=\operatorname{blkdiag}(\overbrace{\mathbf{Q}_{{\rm{d}}}, \ldots, \mathbf{Q}_{{\rm{d}}}}^{\left(N-1\right)}, \overbrace{\mathbf{Q}_{{\rm{a}}}, \ldots, \mathbf{Q}_{{\rm{a}}}}^{N}, \overbrace{\mathbf{Q}_{{\rm{sa}}}, \ldots, \mathbf{Q}_{{\rm{sa}}}}^{N_{\rm{s}}}),
\end{equation}
where $\mathbf{Q}_{{\rm{d}}}\!=\!\operatorname{diag}\left(\delta_{{\rm{d}}}^{2},\left(0.1 \delta_{{\rm{d}}}\right)^{2}\right)$, 
$ \mathbf{Q}_{{\rm{a}}}\!=\!\operatorname{diag}\left(\delta_{{\rm{a}}}^{2}, \delta_{{\rm{a}}}^{2}\right)$, and
$ \mathbf{Q}_{{\rm{sa}}}\!=\!\operatorname{diag}\left((q\delta_{{\rm{a}}})^{2}\!, (q\delta_{{\rm{a}}})^{2}\!, (q\delta_{{\rm{a}}})^{2}\!, (q\delta_{{\rm{a}}})^{2}\right)$; 
$\delta_{\rm{d}}$, $0.1\delta_{\rm{d}}$, and $\delta_{\rm{a}}$ are the standard deviations of the TDOA, FDOA, and AOA measurements of LOS paths, respectively;
$q\delta_{\rm{a}}$ are that of the AOA/AODs of NLOS paths;
scaling factor $q$ accounts for the relative NLOS path noise.
The localization performance is evaluated by the root mean square error (RMSE), e.g., $\text{RMSE}(\mathbf{u})=\sqrt{\sum_{t=1}^{T_{\rm{MC}}}||\mathbf{u}_t-\mathbf{u}^\circ||^2/T_{\rm{MC}}}$, where $\mathbf{u}_t$ is the estimate of $\mathbf{u}^\circ$ at the $t$-th Monte Carlo simulation, and $T_{\rm{MC}}$ is the number of Monte Carlo simulations.
The calculation of CRLBs is given in Appendix \ref{sec:crlb-wls-1}.

\begin{table}[t]
	\vspace{0.8cm}	
	\renewcommand{\arraystretch}{1.5}
	\centering
	\fontsize{8}{8}\selectfont
	\captionsetup{font=small}
	\caption{Locations of the BSs in meters.}\label{tab:bs-location}
	\begin{threeparttable}	
		\begin{tabular}{ccccccc}
			\toprule
			BS index & 1 & 2 & 3 & 4 & 5 & 6 \\
			\hline
			x & 235.504 & 287.504 & 235.504 & 287.504 &  235.504 & 287.504 \\
			y & 389.504 & 389.504 & 489.504 & 489.504 &  589.504 & 589.504 \\
			z & 26 & 32 & 10 & 40 & 14 & 50 \\
			\bottomrule
		\end{tabular}
	\end{threeparttable}
\end{table}

\begin{table}[t]
	\vspace{0.5cm}	
	\renewcommand{\arraystretch}{1.5}
	\centering
	\fontsize{8}{8}\selectfont
	\captionsetup{font=small}
	\caption{Locations of the scatterers in meters, and the index of the corresponding BSs.}\label{tab:scatterer-location}
	\begin{threeparttable}	
		\begin{tabular}{ccccccccccccccccccc}
			\toprule
			& 1 & 2 & 3 & 4 & 5 & 6 & 7 & 8 & 9 & 10 & 11 & 12 & 13 & 14 & 15 & 16 & 17 & 18 \\
			\hline
			x & 247 &240 &280 &255& 270 &266&258& 249 &277 & 265 &276& 238&225 &262& 246&241&251&264  \\
			y & 400& 380& 410& 460& 413& 455&400& 423 &416 & 422&434&462&452&438&444&375&418&444 \\
			z & 6& 10 &12 &15& 7 &10&13& 5 &21 & 14&8&7&9&11&19&4&17&8 \\
			BS index & 1& 2& 3& 4& 5&6 &1 &2 & 3 & 4& 5&6 &1 &2 & 3& 4& 5&6\\
			\bottomrule
		\end{tabular}
	\end{threeparttable}
	\vspace{-0.8cm}
\end{table}

\subsubsection{Performance Evaluation of the First-stage Estimator}
\label{sec:results-wls-1}
\begin{figure}
\vspace{-0.25cm}
\centering
\captionsetup{font=footnotesize}
  \begin{minipage}{0.326\linewidth}
    \centerline{\includegraphics[width=\textwidth]{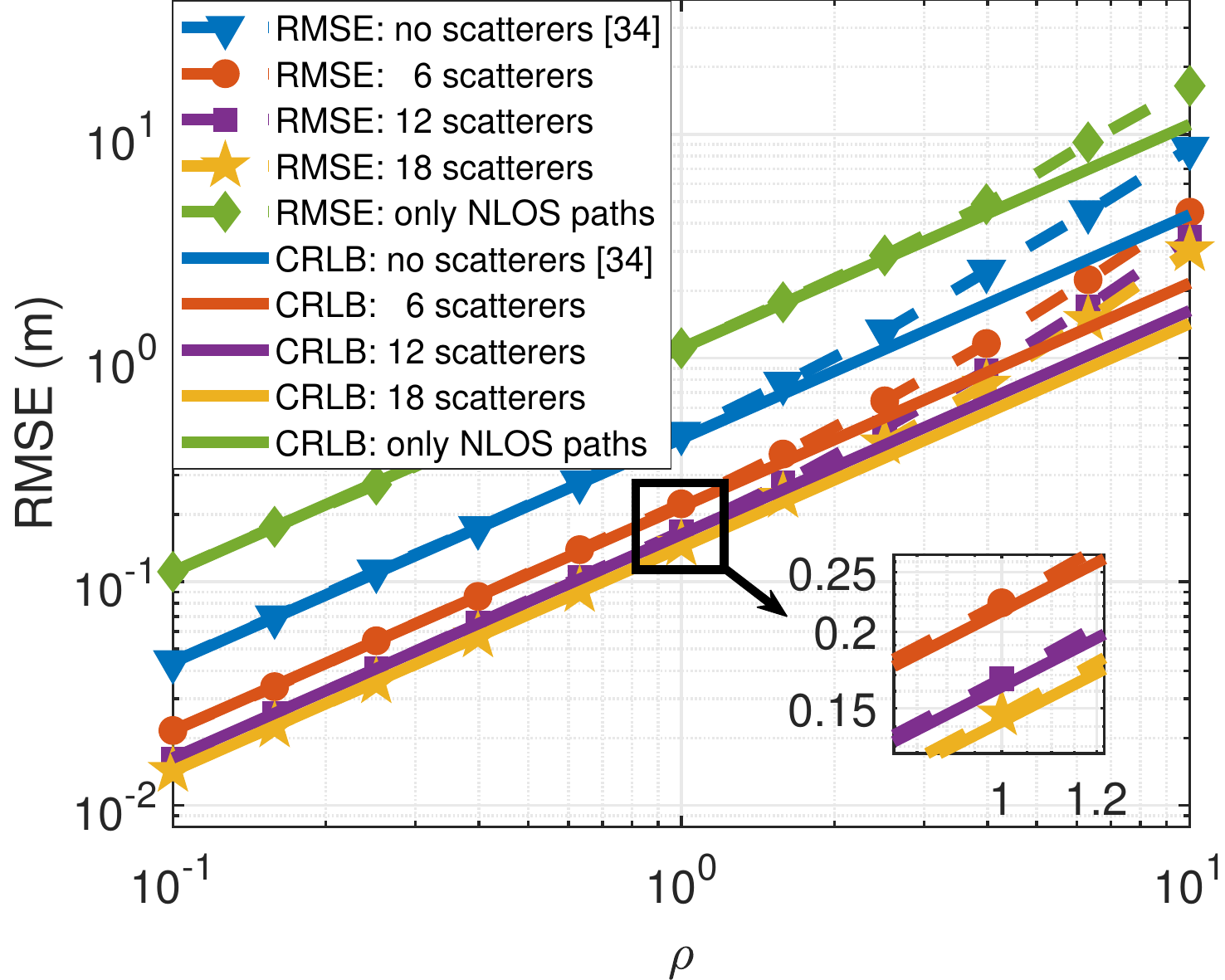}}
    \vspace{-0.1cm}
    \centerline{\footnotesize{(a) UE Location}}
  \end{minipage}
  \hfill
  \begin{minipage}{0.326\linewidth}
    \centerline{\includegraphics[width=\textwidth]{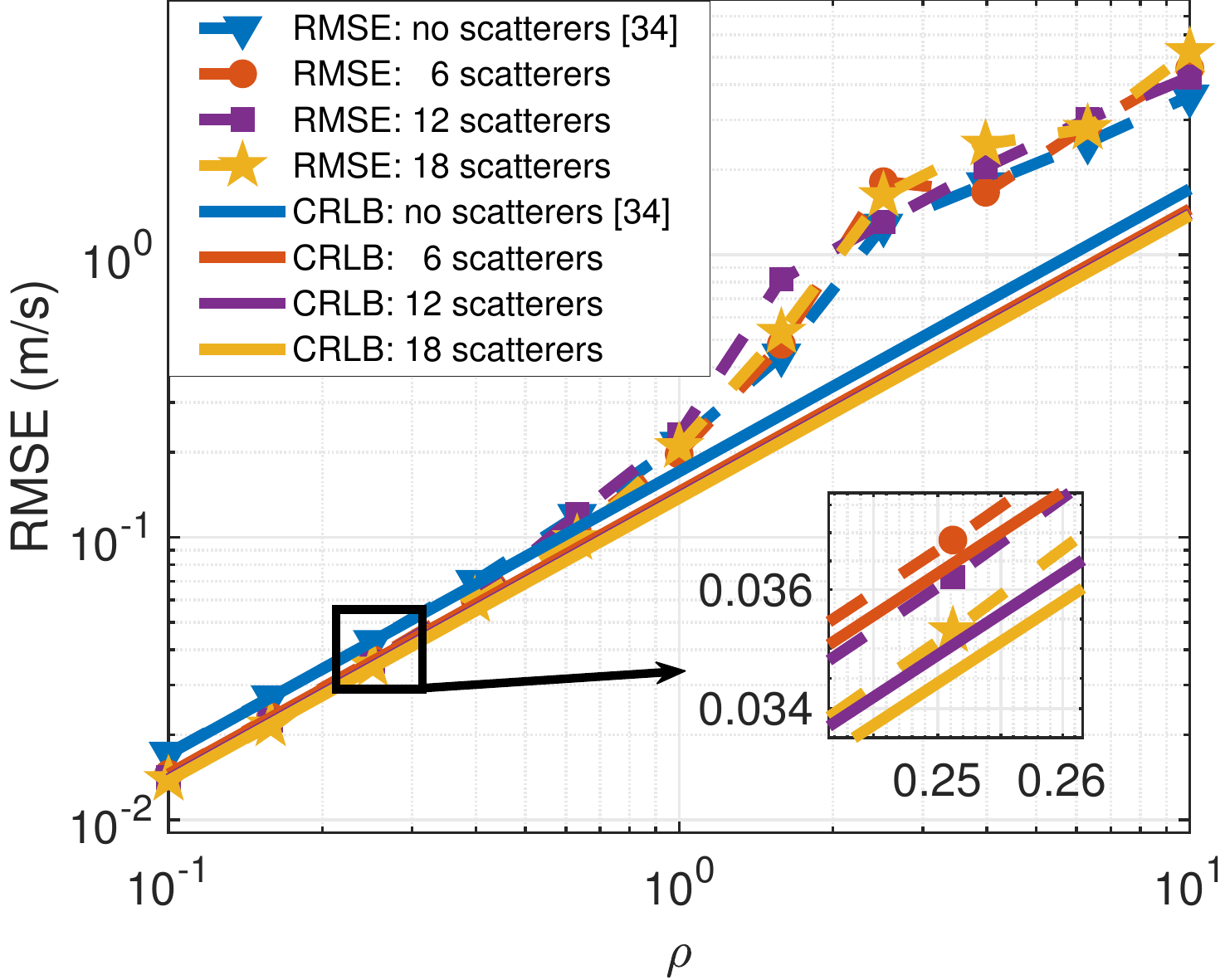}}
    \vspace{-0.1cm}
    \centerline{\footnotesize{(b) UE Velocity}}
  \end{minipage}
  \hfill
  \begin{minipage}{0.326\linewidth}
    \centerline{\includegraphics[width=\textwidth]{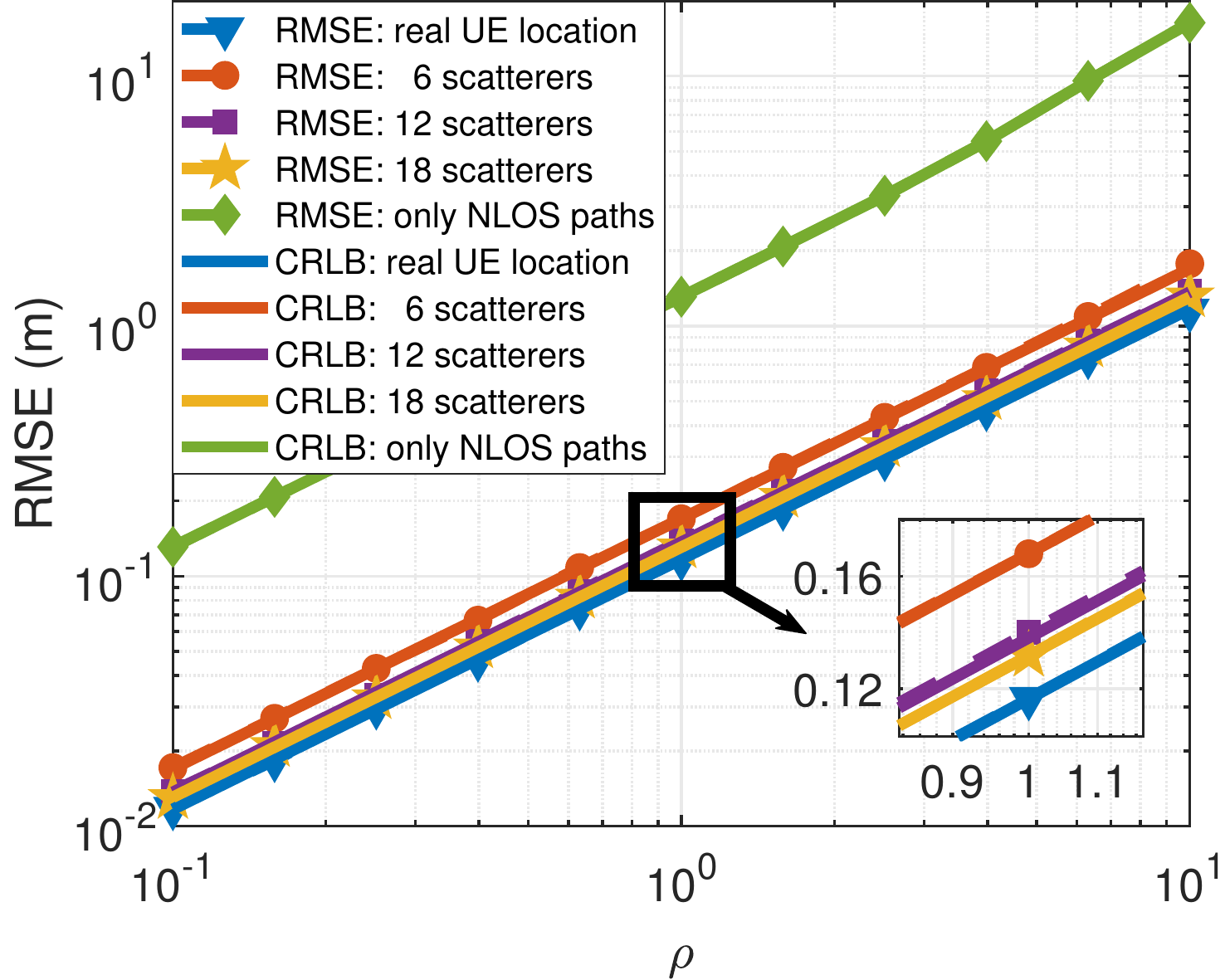}}
    \vspace{-0.1cm}
    \centerline{\footnotesize{(c) Scatterer Location}}
  \end{minipage}
  \vspace{0.2cm}
  \caption{RMSE performance and the corresponding CRLBs of the one-stage estimator with different numbers of scatterers, where six LOS paths are ensured, except the lines labeled by ``only NLOS paths''; the same number of scatterers correspond to each BS.}
  \label{fig:results-wls-1}
  \vspace{-0.1cm}
\end{figure}

We compare the performance of our proposed estimator (using hybrid LOS and NLOS measurements) with the state-of-the-art method in \cite{yang2021model} (using only LOS measurements).
One thousand different UE locations are randomly selected in space $\mathbb{D}\!=\!\left\{[x, y, z]^{T}\!\!\!: 200 \leqslant \!x\! \leqslant 300, 350 \leqslant \!y\! \leqslant 650, 0 \leqslant \!z\! \leqslant 30 \ \text{(unit: m)} \right\}$.
UE velocity is also randomly generated.
For each UE location, we set $T_{\rm{MC}}=1000$.
Moreover, $\delta_{\rm{d}}=0.22\rho$, and $\delta_{\rm{a}}=0.0175\rho$, where $\rho\in\left[0.1,10\right]$ is a scaling factor for the measurement noise. 
NLOS path measurements with small noise are given with $q=0.1$. 
Fig. \ref{fig:results-wls-1} illustrates the simulation results. 
The UE location accuracy is given in Fig. \ref{fig:results-wls-1}(a), where remarkable performance improvements are achieved by utilizing NLOS path measurements.
However, with the increasing number of scatterers, the performance improvements provided by new NLOS paths fade gradually, and 18 scatterers are believed to achieve a satisfactory tradeoff between the localization performance and the number of scatterers.
The CRLBs are achieved under low noise levels, and a slow deviation appears with the increase in measurement noise.
The same phenomena can be found in Fig. \ref{fig:results-wls-1}(b), but the estimation accuracy of UE velocity is only slightly improved, provided that NLOS path measurements are not directly related to UE velocity.
Six scatterers can approach the best estimation accuracy of UE velocity in this scenario.
Furthermore, the estimator simultaneously determines scatterer locations. 
In Fig. \ref{fig:results-wls-1}(c), only the results of the first scatterer are exhibited, whereas those of others are similar.
We take the localization accuracy with the real UE location in \eqref{eq:aoa-aod-wls-eq-nlos} as the floor of the RMSE of scatterer position, which is expected to be attained with an adequate number of scatterers.
The localization results of UE and scatterers with only NLOS paths are also presented in Figs. \ref{fig:results-wls-1}(a) and \ref{fig:results-wls-1}(c), verifying the effectiveness of the estimator under the conditions where LOS paths are all blocked.
However, the localization accuracy is relatively lower than that with hybrid LOS and NLOS path measurements.

\subsubsection{Localization Accuracy vs. NLOS Path Noise}
\begin{figure}
\vspace{-0.25cm}
\centering
\captionsetup{font=footnotesize}
  \begin{minipage}{0.326\linewidth}
    \centerline{\includegraphics[width=\textwidth]{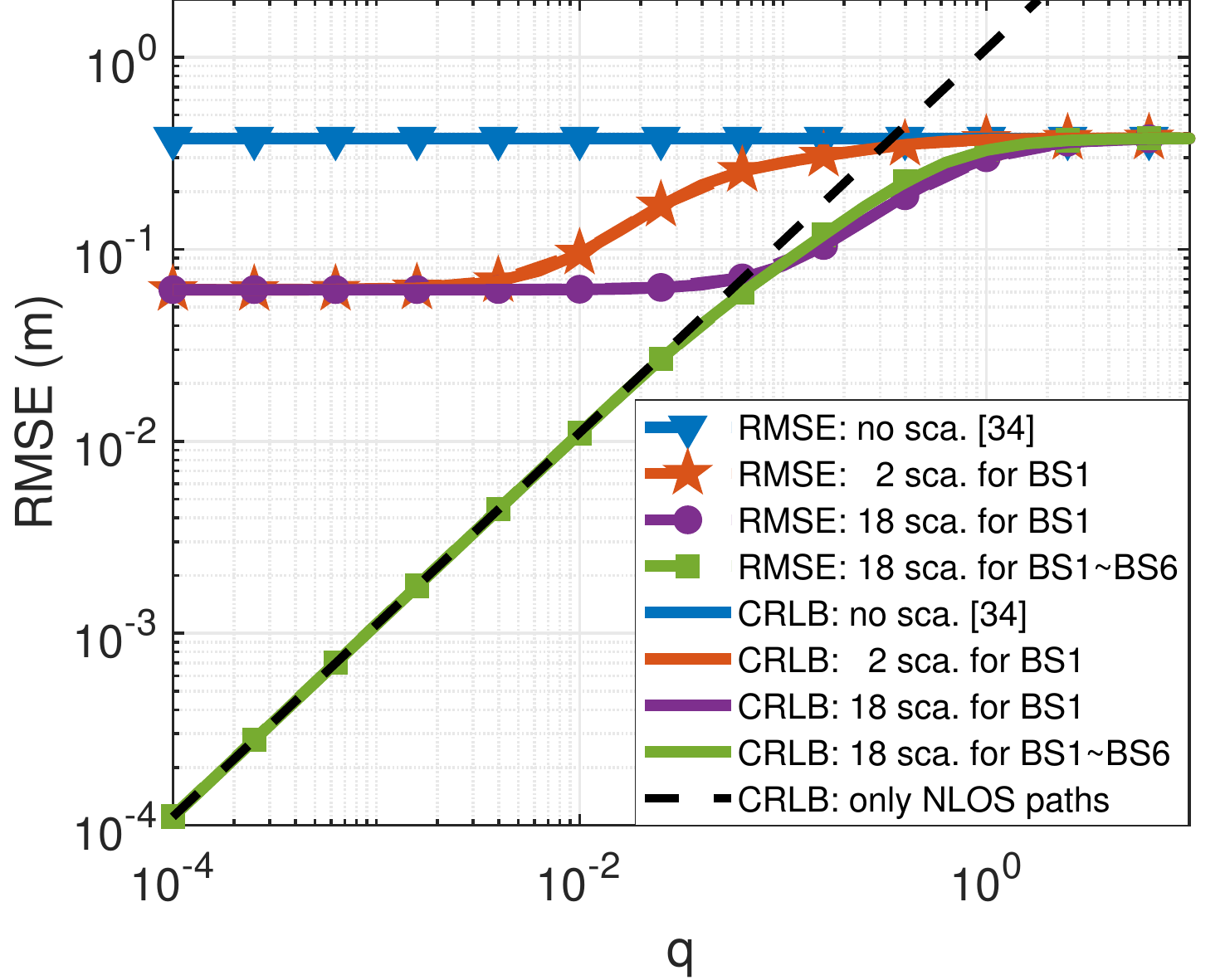}}
    \vspace{-0.1cm}
    \centerline{\footnotesize{(a) UE Location}}
  \end{minipage}
  \hfill
  \begin{minipage}{0.326\linewidth}
    \centerline{\includegraphics[width=\textwidth]{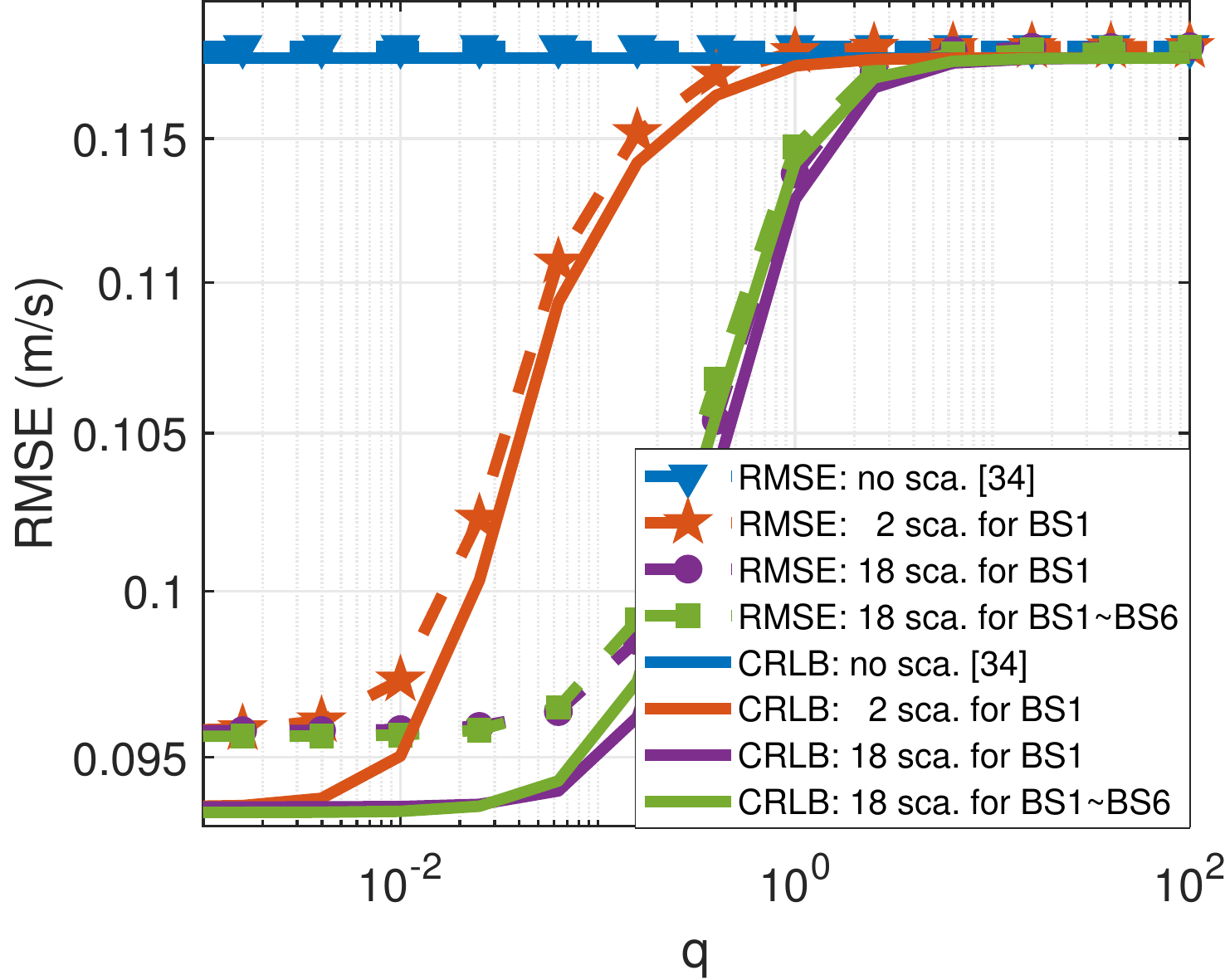}}
    \vspace{-0.1cm}
    \centerline{\footnotesize{(b) UE Velocity}}
  \end{minipage}
  \hfill
  \begin{minipage}{0.326\linewidth}
    \centerline{\includegraphics[width=\textwidth]{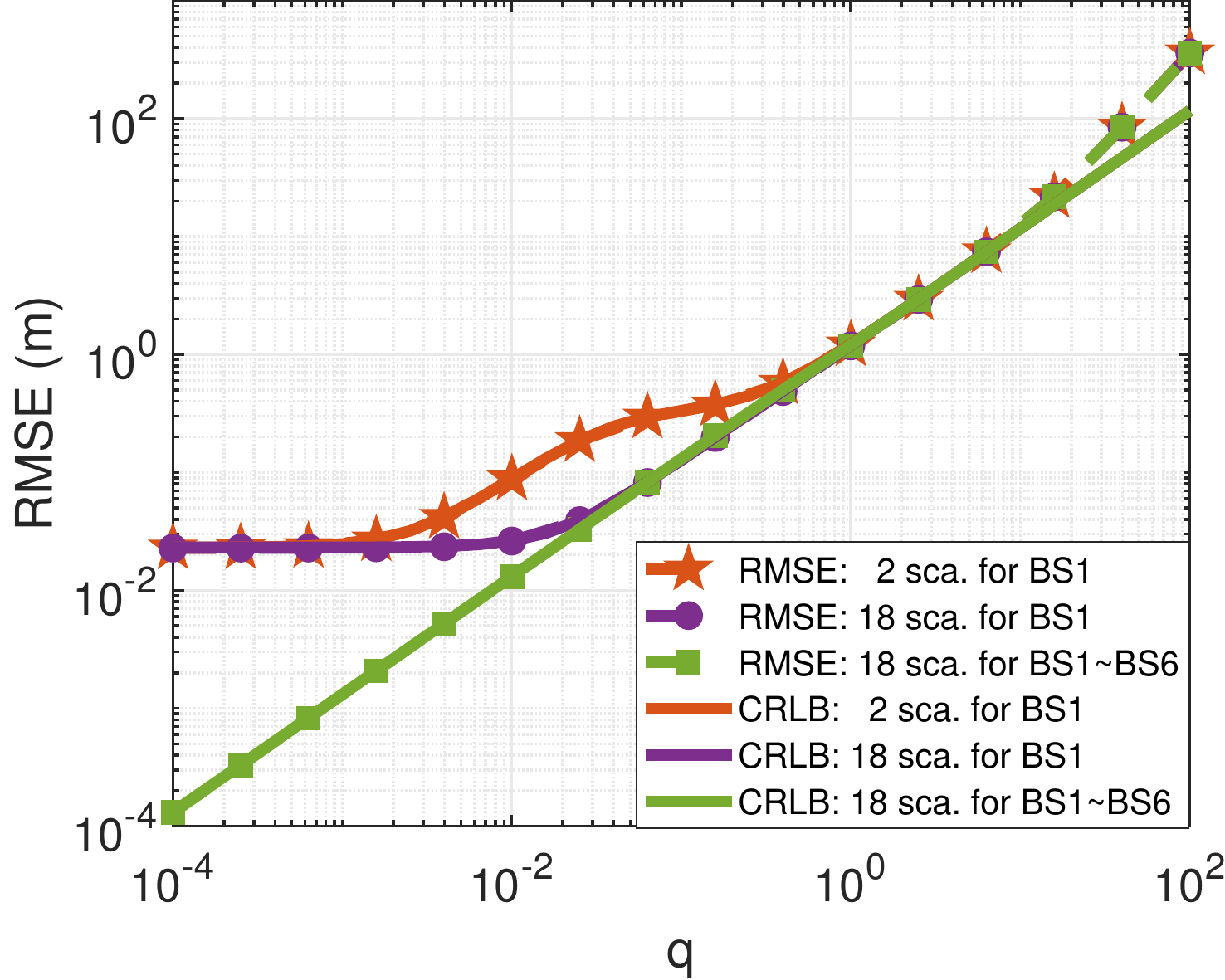}}
    \vspace{-0.1cm}
    \centerline{\footnotesize{(c) Scatterer Location}}
  \end{minipage}
  \vspace{0.2cm}
  \caption{RMSE performance and the corresponding CRLBs of the one-stage estimator with different NLOS path measurement noise (``scatterers'' are simplified as ``sca.'' in the figures).}
  \label{fig:results-wls-1-nlos-noise}
  \vspace{-0.2cm}
\end{figure}

We change the noise scaling factor $q$ to analyze the roles of NLOS paths.
\textcolor{black}{The UE location and velocity are specified by $\mathbf{u}^\circ=[250,450,0]^T\ \text{m}$ and $\dot{\mathbf{u}}^\circ=[-10,2,5]^T\ \text{m/s}$, respectively.}
We set $\delta_{\rm{d}}=0.22\ \text{m}$ and $\delta_{\rm{a}} = 0.0175\ \text{rad}$ with the results shown in Fig. \ref{fig:results-wls-1-nlos-noise}.
Six LOS paths are ensured for most situations, except the dotted line in Fig. \ref{fig:results-wls-1-nlos-noise}(a), which presents the NLOS-only situation.
The RMSEs of location and velocity are positively related to NLOS path measurement noise, whereas some floors and roofs exist.
For UE localization results in Fig. \ref{fig:results-wls-1-nlos-noise}(a), two branches are clarified by determining whether the RMSE has a floor when $q$ is adequately small.
The branch with a floor corresponds to the situation where NLOS paths provide additive but limited information for the estimator and cannot locate UE and scatterers alone.
By contrast, when $q<10^{-2}$, and three or more scatterers link to at least two BSs, the estimator approximately reduces to the special case discussed in Remark \ref{remark:wls-1-special-case}, without floors of RMSE and CRLB.
Moreover, the RMSE of the LOS-only situation in \cite{yang2021model} acts as the roof of that of the proposed estimator, preventing errors caused by inaccurate NLOS path measurements.
Similarly, the results in Fig. \ref{fig:results-wls-1-nlos-noise}(c) also grow into two branches because the localization accuracy of scatterers is strongly dependent on UE location.
However, no roofs exist for the RMSE of scatterer locations because worse NLOS path measurements directly result in worse localization accuracy.
For the RMSEs of UE velocity in Fig. \ref{fig:results-wls-1-nlos-noise}(b), the floor and roof exist simultaneously because the AOA/AODs of NLOS paths are incapable of velocity estimation.
In conclusion, rich scattering environments and NLOS path measurements with small noise are beneficial for the proposed estimator.

\subsubsection{Performance Evaluation of the Second-stage Estimator}
\label{sec:results-wls-2}
\begin{figure}
\centering
\captionsetup{font=footnotesize}
  \begin{minipage}{0.48\linewidth}
    \centerline{\includegraphics[width=\textwidth]{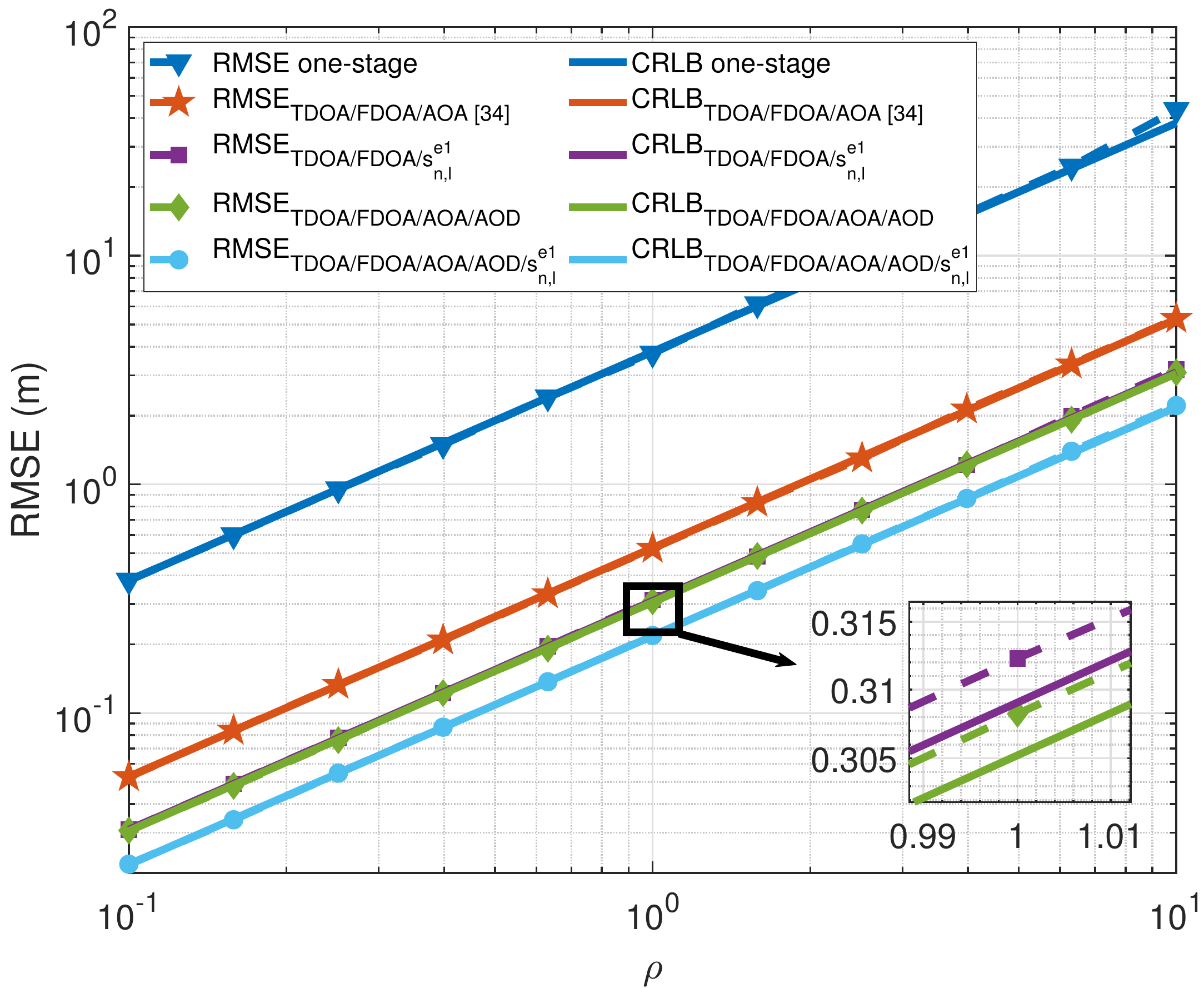}}
    \vspace{-0.1cm}
    \centerline{\footnotesize{(a) Scatterer Location}}
  \end{minipage}
  \hfill
  \begin{minipage}{0.48\linewidth}
    \centerline{\includegraphics[width=\textwidth]{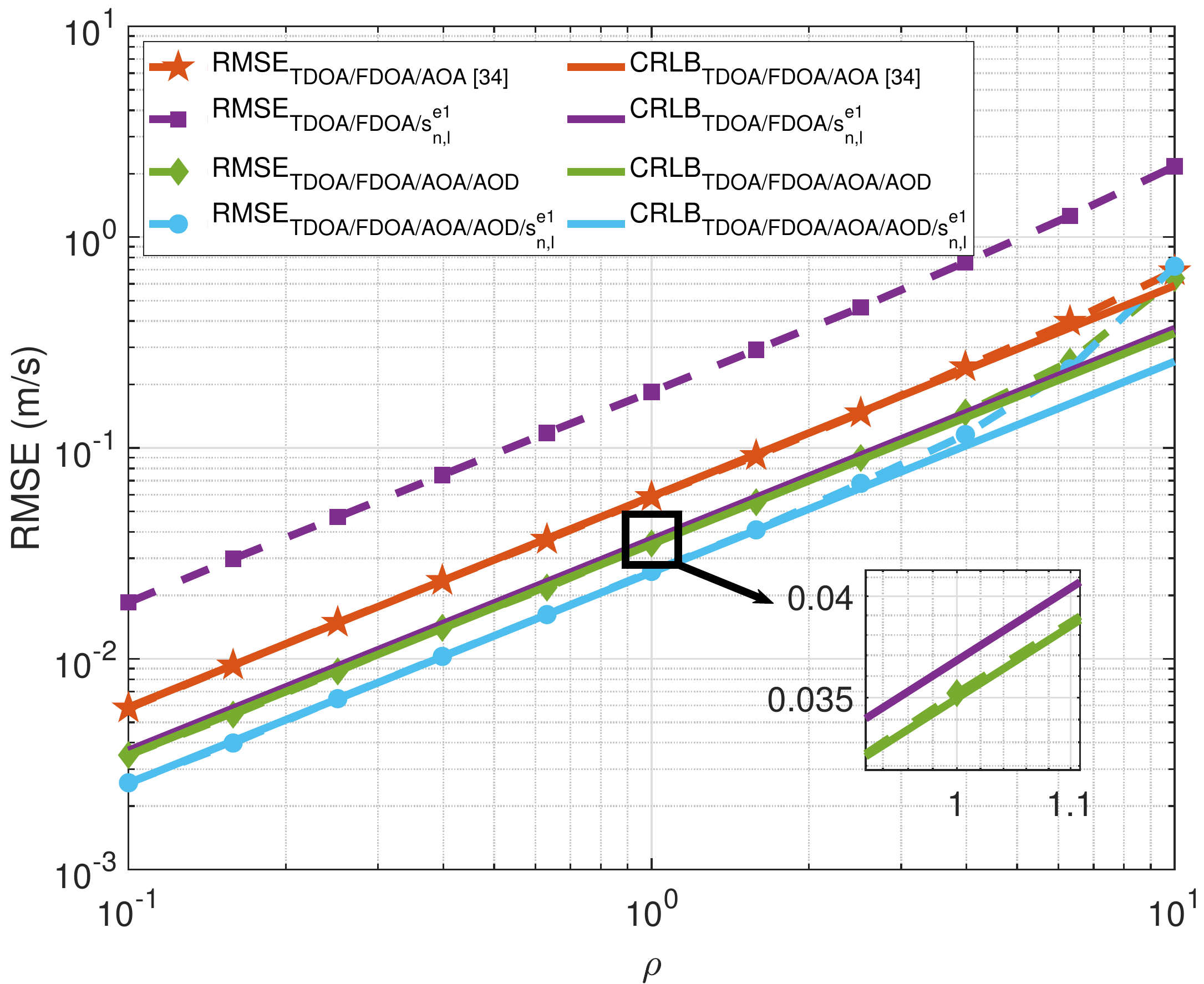}}
    \vspace{-0.1cm}
    \centerline{\footnotesize{(b) Scatterer Velocity}}
  \end{minipage}
  \vspace{0.2cm}
  \caption{Comparison of the RMSEs and CRLBs of the proposed second-stage estimator with the method in \cite{yang2021model}.}
  \label{fig:results-wls-2}
  \vspace{-0.2cm}
\end{figure}

We set $\mathbf{s}_{n,l}^\circ=\left[240,600,-19\right]^T$ m and $\dot{s}_{n,l}^\circ = 5$ m/s to explore the performance of the second-stage estimator.
For comprehensive comparisons, two other simplified versions of the second-stage estimator are discussed in this scenario.
The first one employs the pure channel parameters (TDOA/FDOA/AOA/AOD) for scatterer estimation, to reveal the environmental information provided by $\mathbf{s}_{n,l}^{{\rm{e}1}}$ and $\mathbf{Q}_{n,l}^{\rm{e}1}$. 
The second one employs the TDOA/FDOA parameters and $\mathbf{s}_{n,l}^{{\rm{e}1}}$, as AOA/AOD parameters have been used in the first-stage estimator to derive $\mathbf{s}_{n,l}^{{\rm{e}}1}$, comparing the information provided by $\mathbf{s}_{n,l}^{{\rm{e}}1}$ with that by AOA/AOD parameters.
Simulation results shown in Fig. \ref{fig:results-wls-2} are obtained from $T_{\rm{MC}} = 5000$ independent Monte Carlo simulations, whereas other parameters are the same as that in Sec. \ref{sec:results-wls-1}.
All the discussed solutions can improve scatterer location accuracy and further estimate scatterer velocity.
In Fig. \ref{fig:results-wls-2}(a), the estimator with TDOA/FDOA/AOA/AOD and that with TDOA/FDOA/$\mathbf{s}_{n,l}^{{\rm{e}1}}$ obtain close RMSEs and CRLBs of scatterer location, meaning that $\mathbf{s}_{n,l}^{\rm{e}1}$ provides similar environmental information as AOA/AOD when AOA/AOD are absent.
The estimator with TDOA/FDOA/AOA/AOD/$\mathbf{s}_{n,l}^{{\rm{e}1}}$ achieves the best localization performance, revealing that $\mathbf{s}_{n,l}^{\rm{e}1}$ contains certain environmental information about scatterer location that is not captured by pure NLOS path measurements.
Similar insights are found in the RMSEs of scatterer velocity, as shown in Fig. \ref{fig:results-wls-2}(b).
However, the estimator with TDOA/FDOA/$\mathbf{s}_{n,l}^{{\rm{e}1}}$ has an abnormally bad RMSE performance, which means that part of the information related to scatterer velocity may have been lost in $\mathbf{s}_{n,l}^{\rm{e}1}$.
Finally, the CRLBs can be attained under low noise levels, whereas clear derivations appear in velocity estimation with the gradual increase in measurement noise.

\subsection{RIS-aided Localization}
\label{sec:results-ris-aided}
\begin{figure}
\vspace{-0.45cm}
\centering
\captionsetup{font=footnotesize}
  \begin{minipage}{0.48\linewidth}
    \centerline{\includegraphics[width=0.9\textwidth]{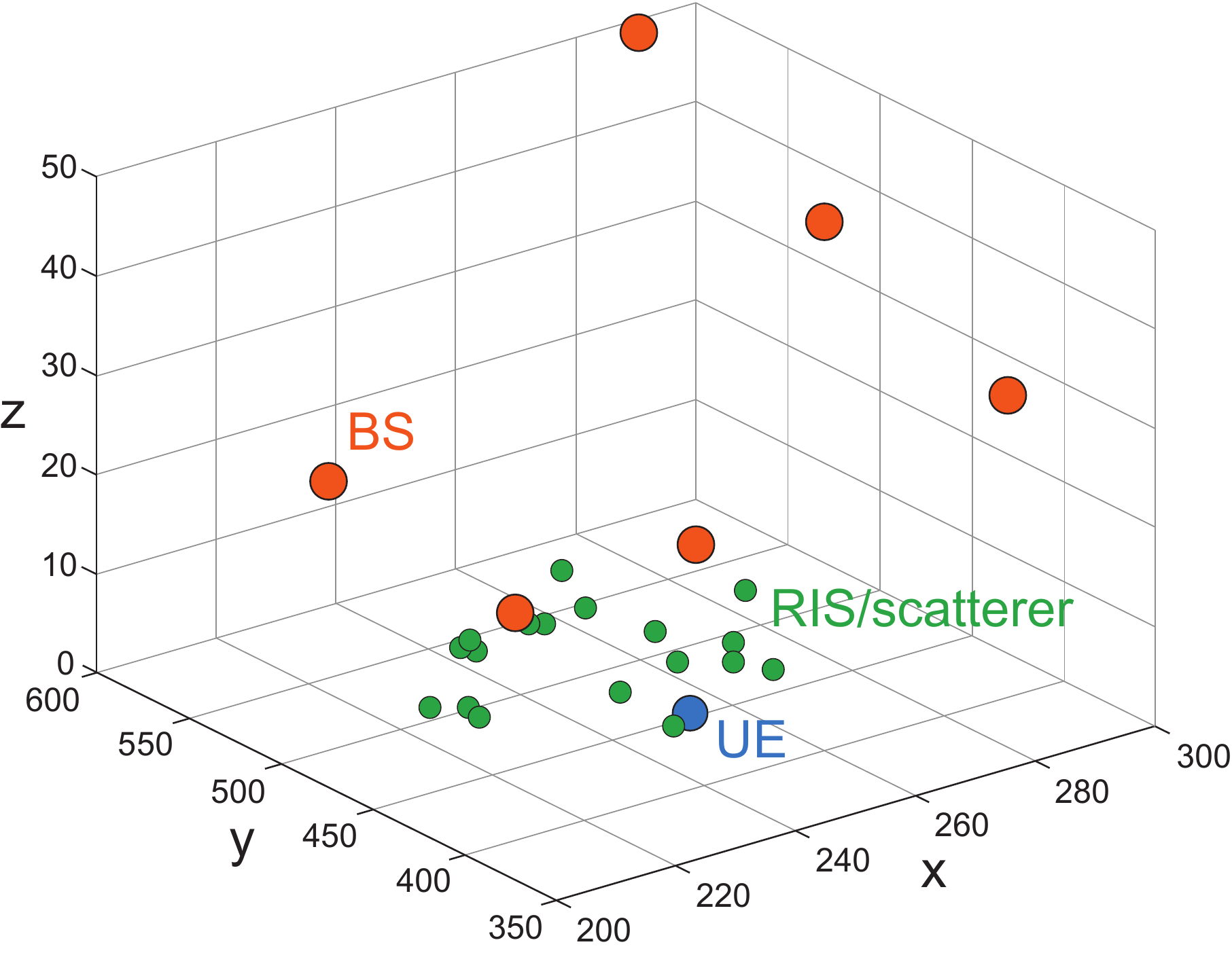}}
    \vspace{-0.05cm}
    \centerline{\footnotesize{(a) Dense Deployment}}
  \end{minipage}
  \begin{minipage}{0.48\linewidth}
    \centerline{\includegraphics[width=0.9\textwidth]{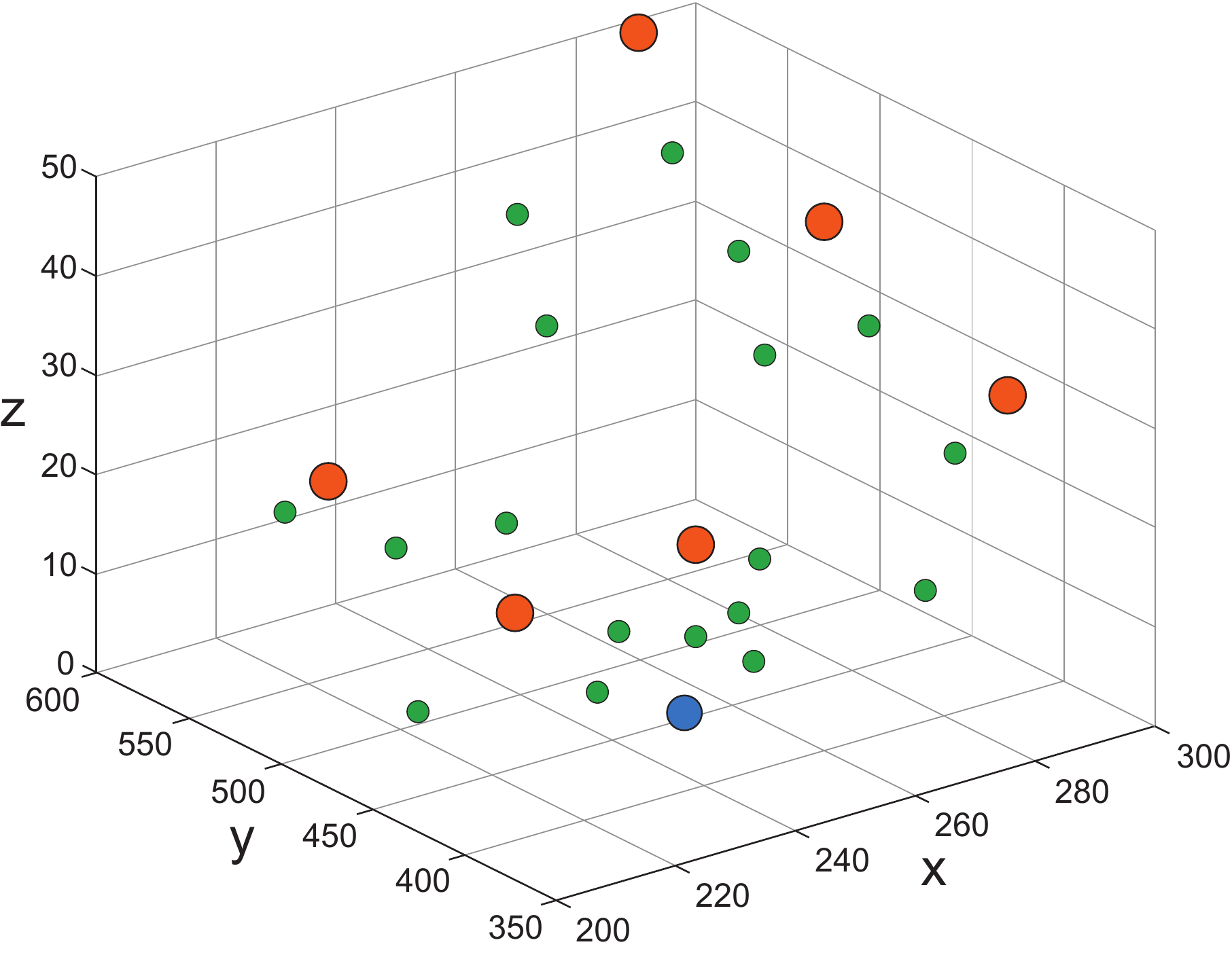}}
    \vspace{-0.05cm}
    \centerline{\footnotesize{(b) Distributed Deployment}}
  \end{minipage}
  \vspace{0.2cm}
  \caption{Two different deployments of RISs/scatterers (Unit: m).}
  \label{fig:results-ris-deployments}
\end{figure}

The proposed estimators in Sec. \ref{sec:wls} have been evaluated to perform well when NLOS paths provide measurements with small noise.
As discussed in Sec. \ref{sec:rcs}, the tunable phase shifts of RIS elements can provide large RCS, which means a strong NLOS path and channel estimates with small noise.
Thus, RIS-aided systems are expected to achieve satisfactory localization performance.
In this subsection, we mainly focus on UE localization, and the CRLB of channel estimation given in Sec. \ref{sec:crlb-channel-estimation} is used as the variance of the corresponding measurements.
Six BSs with known locations and 18 scatterers (or RISs) with unknown locations exist. 
We compare the performances of scatterers and RISs; the dense and distributed deployments of RISs are also discussed, as illustrated in Fig. \ref{fig:results-ris-deployments}.
The former deployment ensures that all RISs are located around UE and the nearest BS to UE, whereas the latter one guarantees that RISs are located around the six BSs in a distributed way.
All locations are chosen randomly, whereas the far-field assumption is ensured.
We assume that the phase shifts of RIS elements have been tuned to maintain the largest RCS in the corresponding incident/scattering angles, as discussed in Remark \ref{remark:new-response1}.

Through the simulation in this subsection,
the center frequency $f=27\  \text{GHz}$, and wavelength $\lambda=11\  \text{mm}$; 
the antenna gain $G=10.96^2$ is responsible for transmitting and receiving antennas; 
the number of subcarriers $H=256$, with the spacing $\Delta f=120\ \text{kHz}$; 
thus, the bandwidth $B=30.72\  \text{MHz}$; 
the number of OFDM symbols $R=256$, and the symbol duration $T_{{\rm{sym}}}=8.3\  \mu s$; 
the size of BS antennas is $15\times 15$, and the UPA of the UE is $5 \times 5$, with the antenna spacing being $\lambda/2$;
the scatterer subpatches and RIS elements are in the sub-wavelength size with $d_{\rm{x}}=d_{\rm{y}}=0.4\lambda$.
The UE is located at $\mathbf{u}^\circ=[250,450,0]^T\ \text{m}$, and the average SNR is 28.5 dB for the LOS paths.

\begin{figure}[t]
	\vspace{0.5cm}
	\centering
	\captionsetup{font=footnotesize}
	\includegraphics[width=0.49\textwidth]{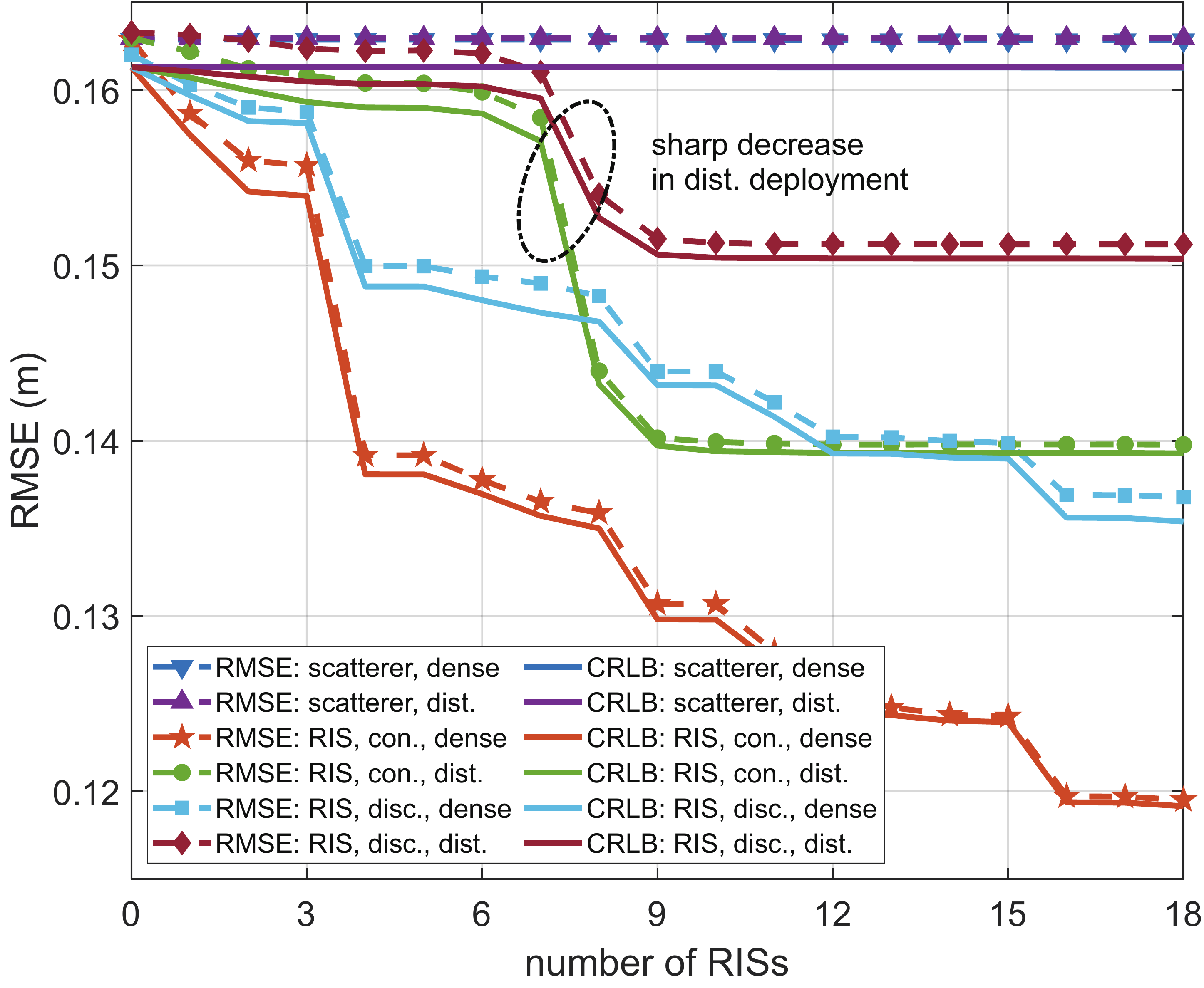}
	\vspace{0.1cm}
	\caption{Comparison of RMSE performance of UE with different numbers of RISs or scatterers in different deployments (abbreviations ``dist.,'' ``con.,'' and ``disc.'' represent ``distributed'' deployment, ``continuous'' phase shifts, and one-bit ``discrete'' phase shifts, respectively).}
	\label{fig:results-rmse-ris-aided-number}
	\vspace{-0.1cm}
\end{figure}

\subsubsection{Localization Accuracy vs. Number of RISs}

We study the localization performance of the first-stage estimator in RIS-aided communication systems.
18 RISs are employed for the estimator in a randomly generated order, and each RIS comprises $100 \times 100$ RIS elements.
Meanwhile, the scatterers of the same shape and size and located at the same positions as RISs are used for comparison. 
From the results illustrated in Fig. \ref{fig:results-rmse-ris-aided-number}, RISs can be found to have a normally good performance in assisting in UE localization because they can maintain large RCSs, leading to strong NLOS paths and channel estimates with small noise.
The RISs with one-bit discrete phase shifts act similarly to those with continuous phase shifts, but the performance improvements in UE localization are relatively small.
By contrast, large RCSs only exist around the specular angles for scatterers, which are rare in practical 3D environments.
Consequently, inaccurate estimates of NLOS paths of traditional scatterers provide little information about UE locations for the estimator.
Therefore, only RISs benefit UE localization in the discussed scenario, whereas scatterer contributions can be neglected (but scatterer locations can be simultaneously estimated).
In addition, densely deployed RISs around UE and the nearest BS to UE have smaller path losses than those in the distributed deployment, thus achieving better location accuracy.
Moreover, the sharp decrease in the RMSEs of the distributed deployments in Fig. \ref{fig:results-rmse-ris-aided-number} precisely corresponds to the RISs located around the nearest BS to UE.
This finding implies that nearer RISs to UE play more important roles in the proposed estimator.
Finally, the CRLBs can be attained in the simulation results, and more RISs achieve tighter bounds.

\begin{figure}[t]
	\vspace{-0.25 cm}
	\begin{minipage}[t]{0.49\textwidth}
		\centering
		\captionsetup{font=footnotesize}
		\includegraphics[width=\textwidth]{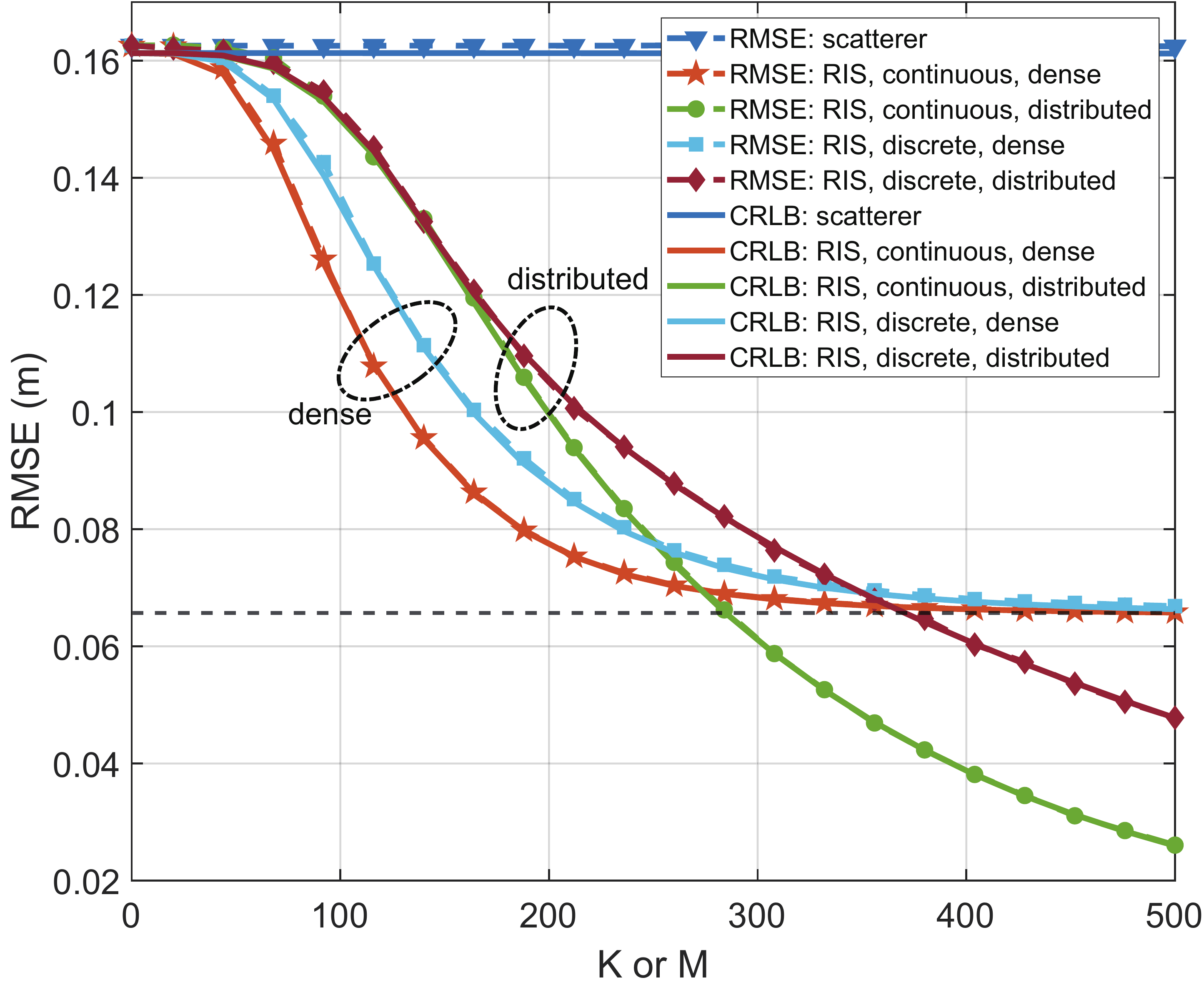}
		\vspace{-0.2cm}
		\caption{RMSE performance in UE location of different sizes and deployments of RISs configured with continuous or discrete optimal phase shifts.}
		\label{fig:results-rmse-ris-aided-size-wls-1}
	\end{minipage}
	\hfill
	\begin{minipage}[t]{0.49\textwidth}
		\centering
		\captionsetup{font=footnotesize}
		\includegraphics[width=\textwidth]{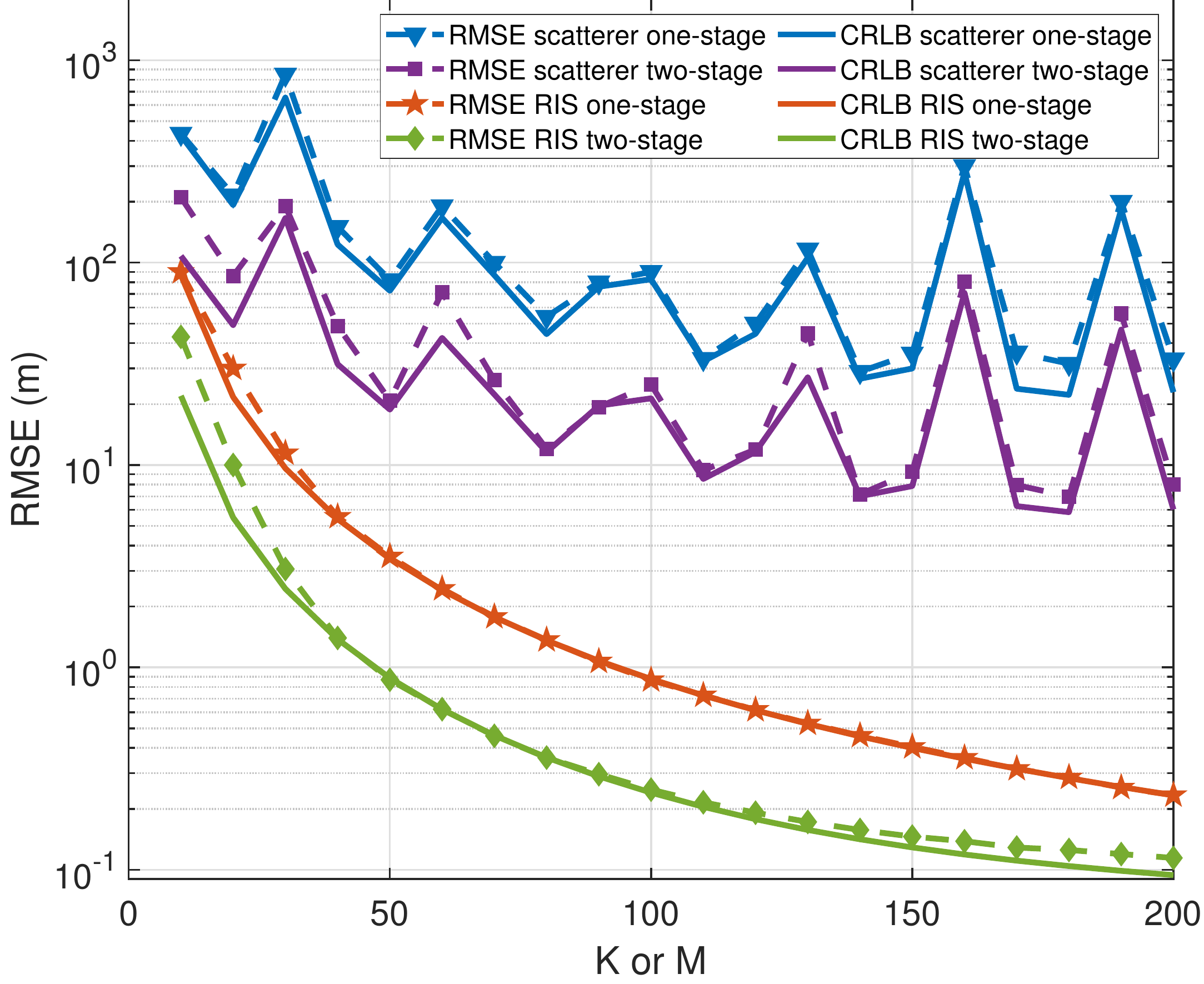}
		\vspace{-0.2cm}
		\caption{RMSE performance in scatterer location of different sizes for first- and second-stage estimators, whereas RISs are configured with optimal continuous phase shifts.}
		\label{fig:results-rmse-ris-aided-size-wls-2}
	\end{minipage}
	\vspace{-0.1 cm}
\end{figure}

\subsubsection{Localization Accuracy vs. RIS Size}

Fig. \ref{fig:results-rmse-ris-aided-size-wls-1} discusses the effects of RIS size on UE localization accuracy, where 18 RISs are considered for the simulation, and $K=M$ is ensured.
Owing to the negligible contributions of scatterers, we no longer distinguish their deployments.
As the results illustrate, larger RISs are believed to obtain higher location accuracy.
In the dense deployment, the RMSE floor in Fig. \ref{fig:results-wls-1-nlos-noise}(a) is attainable with adequate RIS elements and broken in the distributed deployment because the RISs link to different BSs.
The performance of RISs with discrete phase shifts follows the same trend as that with continuous phase shifts, but the location accuracy is slightly degraded.
Notably, the simulation results may lack sufficient accuracy when an RIS is extremely large, given the necessity to consider near-field effects.
Moreover, we verify the effectiveness of our proposed second-stage estimator with continuous phase shifts in Fig. \ref{fig:results-rmse-ris-aided-size-wls-2}.
RIS localization is much more accurate than scatterer localization, and the second-stage estimator substantially improves the accuracy.
Larger RISs provide higher RCSs and measurements with smaller noise, resulting in locations with higher accuracy.
However, as discussed in Fig. \ref{fig:simu-rcs-vs-size}, the RCS of the scatterer grows to the size but with obvious fluctuations.
Accordingly, the RMSEs and CRLBs in Fig. \ref{fig:results-rmse-ris-aided-size-wls-2} decrease up and down in the same manner as the RCS, revealing the different performances of scatterers and RISs in environment sensing.

\section{Conclusion}
\label{sec:conclusion}
In this study, we explore the joint localization and environment sensing problem in RIS-assisted mmWave communication systems. 
We first compare the characteristics of scatterers and RISs in terms of RCS, deriving the far-field RCS of RISs. 
We find that the RIS with identical phase shifts has the same scattering properties as traditional scatterers, whereas optimal phase shifts can result in large RCS and strong NLOS paths. 
Next, we introduce the one-stage WLS estimator that leverages LOS and NLOS paths to perform joint localization and environment sensing. 
This estimator is also tested in scenarios where only NLOS paths are available. 
To further enhance environment sensing accuracy, the second-stage estimator is proposed, utilizing the outputs of the first-stage estimator and NLOS path measurements. 
Simulation results show that the proposed estimators perform well in rich scattering environments, outperform the state-of-the-art methods, and approach the CRLBs with low measurement noise. 
The results also demonstrate the superiority of using NLOS paths in the first-stage estimator for UE localization and in the second-stage estimator for environment sensing. 
Finally, we compare the localization performances of RISs and scatterers and analyze the impacts of various RIS parameters on these performances.

\begin{appendices}
\section{}\label{appendix:ris-prove}
In this section, we give the proof of Theorem \ref{theorem:rcs-ris}.
First, we donate $P_{k, m}^{\rm{in}}$ as the power of the incident waves into the $(k,m)$-th RIS element. 
Taking $Z_0$ as the characteristic impedance of the air, the electric field intensity of the incident signal can be given by
\vspace{-0.2cm}
\begin{equation}\label{eq:e-field-in}
E_{k, m}^{\rm{in}}=\sqrt{\frac{2 Z_{0} P_{k, m}^{\rm{in}}}{d_{\rm{x}} d_{\rm{y}}}} e^{\frac{-j 2 \pi \mathcal{d}_{k, m}^{\rm{t}}}{\lambda}}.
\vspace{-0.2cm}
\end{equation}
According to the law of energy conservation, we have $P_{k, m}^{\rm{in}}\left|\Gamma_{k, m}\right|^{2}=P_{k, m}^{\rm{r e f l e c t}}$, 
where $\Gamma_{k, m}=A_{} e^{-j \xi_{k, m}}$ is the reflection coefficient of the $(k,m)$-th RIS element. 
The scattering gain of the RIS element is given by $G_{\rm{RIS}} = \frac{4\pi d_{\rm{x}}d_{\rm{y}}}{  \lambda ^2}$, 
and the total power of the reflected signal at the receiver is given as $P_{k, m}^{{\rm{r}}}=\frac{G_{{\rm{RIS}}} P_{k, m}^{{\rm{r e f l e c t}}}}{4 \pi d_{k, m}^{{\rm{r}}}{ }^{2}}  F\left(\vartheta_{k,m}^{{\rm{r}}}, \varphi_{k,m}^{{\rm{r}}}\right) F\left(\vartheta_{k,m}^{{\rm{t}}}, \varphi_{k,m}^{{\rm{t}}}\right)  A_{{\rm{r}}}$, where $A_{\rm{r}}$ is the physical aperture of the receiving antenna.
Thus, the electric field intensity at the receiver of the signal reflected by the $(k,m)$-th RIS element is
\vspace{-0.3cm}
\begin{equation}\label{eq:e-field-out-element}
E_{k, m}^{{\rm{r}}}=\sqrt{\frac{2 Z_{0} P_{k, m}^{{\rm{r}}}}{A_{{\rm{r}}}}} e^{\frac{-j2\pi (d_{k,m}^{\rm{t}} + d_{k,m}^{\rm{r}})}{\lambda}} e^{-j\xi_{k,m}}.
\vspace{-0.3cm}
\end{equation}
Second, under the far-field assumption, we have $|E^{\rm{in}}|=|E_{k, m}^{\rm{in}}|$ provided that the area does not affect the absolute value of the electric field intensity of the incident signal.
The electric field intensity at the receiver of the signal reflected by the whole RIS is
\vspace{-0.1cm}
\begin{equation}\label{eq:e-field-out-whole}
E_{\rm{r}} = \sum_{k=1}^K\sum_{m=1}^M E^{\rm{r}}_{k,m}e^{\frac{-j2\pi (d_{k,m}^{\rm{t}} + d_{k,m}^{\rm{r}})}{\lambda}} e^{-j\xi_{k,m}}.
\vspace{-0.1cm}
\end{equation}
With the incident and reflected electric field intensity in Equations \eqref{eq:e-field-in}, \eqref{eq:e-field-out-element}, and \eqref{eq:e-field-out-whole}, we adopt the far-field assumption as \eqref{eq:rcs-scatterer-summation-subpatch-farfield} and derive the RCSs of the RIS element and the whole RIS in Theorem \ref{theorem:rcs-ris}.

\vspace{-0.4cm}
\section{}\label{appendix:wls-equations}
\vspace{-0.1cm}
This section gives the specific forms of the pseudo linear equations and the approximating matrices omitted in Sec. \ref{sec:wls-1-stage} and \ref{sec:wls-2-stage}.
In Sec. \ref{sec:wls-1-stage}, the equations related to TDOA/FDOAs of LOS paths are given as
{\setlength\abovedisplayskip{2pt}
\setlength\belowdisplayskip{4pt}
\begin{align}
\left(r_{n 1,0}^{\circ}\right)^{2}-2 r_{n 1,0}^{\circ} \mathbf{d}^{{\rm{r}}\circ T}_{1,0} \mathbf{b}_{1}-\mathbf{b}_{n}^{T} \mathbf{b}_{n}+\mathbf{b}_{1}^{T} \mathbf{b}_{1}=2\left[\left(\mathbf{b}_{1}-\mathbf{b}_{n}\right)^{T}-r_{n 1,0}^{\circ} \mathbf{d}^{{\rm{r}}\circ T}_{1,0}\right] \mathbf{u}^{\circ},\label{eq:tdoa-wls-eq} \\
\dot{r}_{n 1,0}^{\circ} r_{n 1,0}^{\circ}-\dot{r}_{n 1,0}^{\circ} \mathbf{d}^{{\rm{r}}\circ T}_{1,0} \mathbf{b}_{1}=-\dot{r}_{n 1,0}^{\circ} \mathbf{d}^{{\rm{r}}\circ T}_{1,0} \mathbf{u}^{\circ}+\left[\left(\mathbf{b}_{1}-\mathbf{b}_{n}\right)^{T}-r_{n 1,0}^{\circ} \mathbf{d}^{{\rm{r}}\circ T}_{1,0}\right] \dot{\mathbf{u}}^{\circ},\label{eq:fdoa-wls-eq}
\end{align}
}
\hspace{-0.1cm}where $n = 2, 3, \ldots, N$, and $\mathbf{d}\left(\phi_{1,0}^{\rm{r}\circ},\theta_{1,0}^{\rm{r}\circ}\right)$ are simplified as $\mathbf{d}^{{\rm{r}}\circ}_{1,0}$. 
Similarly, $\mathbf{c}\left(\phi_{n,l}^{\star\circ},\theta_{n,l}^{\rm{r}\circ}\right)$ and $\mathbf{v}\left(\phi_{n,l}^{\star\circ},\theta_{n,l}^{\rm{r}\circ}\right)$ are simplified as $\mathbf{c}^{\star\circ}_{n,l}$ and $\mathbf{v}^{\star\circ}_{n,l}$, where $\star \in \{{\rm{r},\rm{t}}\}$.
Then, the equations related to the AOAs of LOS paths and AOA/AODs of NLOS paths are given as
\vspace{-0.2cm}
\begin{equation}
\mathbf{c}_{n,0}^{{\rm{r}}\circ T} \mathbf{b}_{n}=\mathbf{c}_{n,0}^{{\rm{r}}\circ T} \mathbf{u}^{\circ}, \ \ \mathbf{v}_{n,0}^{{\rm{r}}\circ T} \mathbf{b}_{n}=\mathbf{v}_{n,0}^{{\rm{r}}\circ T} \mathbf{u}^{\circ},\label{eq:aoa-wls-eq-los}
\vspace{-0.3cm}
\end{equation}
\vspace{-0.6cm}
\begin{equation}
\mathbf{c}_{n,l}^{{\rm{r}}\circ T} \mathbf{b}_{n}=\mathbf{c}_{n,l}^{{\rm{r}}\circ T} \mathbf{s}^{\circ }_{n,l}, \ \ \mathbf{v}_{n,l}^{{\rm{r}}\circ T} \mathbf{b}_{n}=\mathbf{v}_{n,l}^{{\rm{r}}\circ T} \mathbf{s}^{\circ }_{n,l}, \ \ \mathbf{c}_{n, l}^{{\rm{t}} \circ T} \mathbf{s}_{n, l}^{\circ}=\mathbf{c}_{n, l}^{{\rm{t}} \circ T} \mathbf{u}^{\circ}, \ \  \mathbf{v}_{n, l}^{{\rm{t}} \circ T} \mathbf{s}_{n, l}^{\circ}=\mathbf{v}_{n, l}^{{\rm{t}} \circ T} \mathbf{u}^{\circ} ,  \label{eq:aoa-aod-wls-eq-nlos}
\vspace{-0.2cm}
\end{equation}
where $n = 1, 2, \ldots, N$, and $l = 1, 2, \ldots, L_n$.
Matrix $\mathbf{B}$ in \eqref{eq:wls-1-error-vector-approx} is given as $\mathbf{B} = \text{blkdiag}(\mathbf{B}_{\rm{s}},\mathbf{B}_{\rm{u}})$, where
{\setlength\abovedisplayskip{6pt}
\setlength\belowdisplayskip{6pt}
\begin{equation}\label{eq:matrix-b-wls-1}
\setlength{\arraycolsep}{2pt}
\begin{aligned}
& \mathbf{B}_{\rm{u}}=\left[\begin{array}{cc}
\mathbf{B}_{1} & \mathbf{B}_{2} \\[-4pt]
\mathbf{O} & \mathbf{B}_{3}
\end{array}\right], \mathbf{B}_{1}=\text{blkdiag}\left(\left[\begin{array}{cc}
2 r_{2, 0}^{\circ} & 0 \\[-4pt]
\dot{r}_{2, 0}^{\circ} & r_{2, 0}^{\circ}
\end{array}\right], \ldots,\left[\begin{array}{cc}
2 r_{N, 0}^{\circ} & 0 \\[-4pt]
\dot{r}_{N, 0}^{\circ} & r_{N, 0}^{\circ}
\end{array}\right]\right) ,
\mathbf{B}_{2}=\left[\begin{array}{ll}
\mathbf{B}_{21} & \mathbf{O}\end{array}\right] ,\\[-2pt]
&\mathbf{B}_{21}\!\!=\!\!\left[\!\begin{array}{lllllllll}
0 & 0 ;& a_{2} & b_{2} ;& \ldots ; & 0 & 0 ;& a_{N} & b_{N}\end{array}\!\right]\!, \ 
\mathbf{B}_{3}\!=\!\text{diag}\left(r_{1, 0}^{\circ} \cos \theta_{1,0}^{{\rm{r}}\circ}, r_{1, 0}^{\circ}, \ldots, r_{N, 0}^{\circ} \cos \theta_{N,0}^{{\rm{r}}\circ}, r_{N, 0}^{\circ}\right)\!,\\[-2pt]
&\mathbf{B}_{{\rm{s}}}  = \text{blkdiag}\left(\mathbf{B}_{1,1}, \ldots, \mathbf{B}_{n,l}, \ldots, \mathbf{B}_{N,L_{N}}\right),\mathbf{B}_{n,l}  = \text{diag}\left(d_{n,l}^{{\rm{r}}\circ}\cos\theta_{n,l}^{{\rm{r}}\circ}, d_{n,l}^{{\rm{r}}\circ}, d_{n,l}^{{\rm{t}}\circ}\cos\theta_{n,l}^{{\rm{t}}\circ}, d_{n,l}^{{\rm{t}}\circ}\right).
\end{aligned}
\end{equation}}
\hspace{-0.15cm}In Sec. \ref{sec:wls-2-stage}, the first two equations in \eqref{eq:wls-2-collected-matrix} related to TDOA/FDOAs of NLOS paths are given as
{\setlength\abovedisplayskip{4pt}
\setlength\belowdisplayskip{-0pt}
\begin{align}
&\left(r_{n, l}^{\circ}\right)^{2}+2 r_{n, l}^{\circ} \mathbf{d}_{n,l}^{\rm{r}\circ} \mathbf{b}_{n}-\mathbf{u}^{\circ T} \mathbf{u}^{\circ}+\mathbf{b}_{n}^{T} \mathbf{b}_{n} = 2\left(\mathbf{b}_{n}-\mathbf{u}^{\circ}+r_{n, l}^{\circ} \mathbf{d}_{n,l}^{\rm{r}\circ}\right)^{T} \mathbf{s}_{n,l}^\circ, \label{eq:detailed-wls-2-eq1}\\[-0pt]
&r_{n, l}^{\circ} \dot{r}_{n, l}^{\circ}\!+\!\dot{r}_{n, l}^{\circ} \mathbf{d}_{n,l}^{\rm{r}\circ} \mathbf{b}_{n}\!-\!\dot{\mathbf{u}}^{\circ T} \mathbf{u}^{\circ} = \left(\dot{r}_{n, l}^{\circ} \mathbf{d}_{n,l}^{\rm{r}\circ}\!-\!\dot{\mathbf{u}}^{\circ}\right)^{T} \mathbf{s}_{n,l}^\circ + \left(r_{n, l}^{\circ} \mathbf{d}_{n,l}^{\rm{r}\circ}\!+\!\mathbf{b}_{n}\!-\!\mathbf{u}^{\circ}\right)^{T} \dot{\mathbf{s}}_{n,l}^\circ. \label{eq:detailed-wls-2-eq2}
\end{align}}

\vspace{-0.8cm}
\section{}\label{sec:crlb-wls-1}
In this section, we derive the CRLB of the proposed one-stage WLS estimator in Sec. \ref{sec:wls-1-stage}.
According to \cite{kay1993fundamentals}, the CRLB of $\mathbf{x}^{\circ}$ for the Gaussian noise model can be given as
\vspace{-0.2cm}
\begin{equation}
\label{eq:crlb-final}
\mathbf{C R L B}\left(\mathbf{x}^{\circ}\right)=\left(\mathbf{D}^{\circ T} \mathbf{Q}^{-1} \mathbf{D}^\circ\right)^{-1},
\vspace{-0.1cm}
\end{equation}
where $\mathbf{D}^\circ=\frac{\partial \mathbf{m}^{\circ}}{\partial \mathbf{x}^{\circ T}} = \left[(\frac{\partial \mathbf{m}_1^{\circ}}{\partial \mathbf{x}^{\circ T}})^T, (\frac{\partial \mathbf{m}_2^{\circ}}{\partial \mathbf{x}^{\circ T}})^T\right]^T$, and $\frac{\partial \mathbf{m}_2^{\circ}}{\partial \mathbf{x}^{\circ T}}=\left[(\frac{\partial \mathbf{m}_{2,1}^{\circ}}{\partial \mathbf{x}^{\circ T}})^T, (\frac{\partial \mathbf{m}_{2,2}^{\circ}}{\partial \mathbf{x}^{\circ T}})^T, \ldots, (\frac{\partial \mathbf{m}_{2,N}^{\circ}}{\partial \mathbf{x}^{\circ T}})^T\right]^T$;
the covariance matrix $\mathbf{Q}$ is composed of the CRLBs of channel parameters given in Sec. \ref{sec:crlb-channel-estimation}.
Here, the partial derivatives of the vectors are composed of the partial derivatives of the channel parameters to the unknown parameters $\mathbf{x}^\circ$. 
Take the parameter $\phi_{n, l}^{{\rm{r}}\circ}$ as an example, we have $\frac{\partial \phi_{n, l}^{{\rm{r}}\circ}}{\partial \mathbf{x}^{\circ T}}=\left[\frac{\partial \phi_{n, l}^{{\rm{r}}\circ}}{\partial \mathbf{u}^{\circ T}}, \frac{\partial \phi_{n, l}^{{\rm{r}}\circ}}{\partial \dot{\mathbf{u}}^{\circ T}}, \frac{\partial \phi_{n, l}^{{\rm{r}}\circ}}{\partial \mathbf{s}^{\circ T}}\right]$, 
and the same formats can be applied to other components.

Then, we list the partial derivatives of the channel parameters used in the estimator. According to the Equations \eqref{eq:tdoa}-\eqref{eq:los-aoa}, we have
\vspace{-0.0cm}
\begin{equation}
\begin{aligned}
\frac{\partial r_{i 1,0}^{\circ}}{\partial \mathbf{u}^{\circ T}}=&\frac{\left(\mathbf{u}^{\circ}-\mathbf{b}_{i}\right)^{T}}{r_{i,0}^{\circ}}-\frac{\left(\mathbf{u}^{\circ}-\mathbf{b}_{1}\right)^{T}}{r_{1,0}^{\circ}}, \quad \frac{\partial r_{i 1,0}^{\circ}}{\partial \dot{\mathbf{u}}^{\circ T}}=\mathbf{0}, \quad \frac{\partial r_{i 1,0}^{\circ}}{\partial \mathbf{s}_{n,l}^{\circ T}}=\mathbf{0}, \\[-4pt]
\frac{\partial \dot{r}_{i 1,0}^{\circ}}{\partial \mathbf{u}^{\circ T}}=&\ \frac{\dot{r}_{1,0}^{\circ}\left(\mathbf{u}^{\circ}-\mathbf{b}_{1}\right)^{T}}{\left(r_{1,0}^{\circ}\right)^{2}}-\frac{\dot{r}_{i,0}^{\circ}\left(\mathbf{u}^{\circ}-\mathbf{b}_{i}\right)^{T}}{\left(r_{i,0}^{\circ}\right)^{2}}+\frac{\dot{\mathbf{u}}^{\circ T}}{r_{i,0}^{\circ}}-\frac{\dot{\mathbf{u}}^{\circ T}}{r_{1,0}^{\circ}}, \quad \frac{\partial \dot{r}_{i 1,0}^{\circ}}{\partial \mathbf{s}_{n,l}^{\circ T}}=\mathbf{0},\\[-4pt]
\quad \frac{\partial \dot{r}_{i 1,0}^{\circ}}{\partial \dot{\mathbf{u}}^{\circ T}}=&\ \frac{\left(\mathbf{u}^{\circ}-\mathbf{b}_{i}\right)^{T}}{r_{i,0}^{\circ}}-\frac{\left(\mathbf{u}^{\circ}-\mathbf{b}_{1}\right)^{T}}{r_{1,0}^{\circ}}, \quad \frac{\partial \phi_{j,0}^{{\rm{r}}\circ}}{\partial \mathbf{u}^{\circ T}}=\frac{\mathbf{c}_{j,0}^{{\rm{r}}\circ T}}{r_{j,0}^{\circ} \cos \theta_{j,0}^{{\rm{r}}\circ}}, \quad \frac{\partial \theta_{j,0}^{{\rm{r}}\circ}}{\partial \mathbf{u}^{\circ T}}=\frac{\mathbf{v}_{j,0}^{{\rm{r}}\circ T}}{r_{j,0}^{\circ}},
\end{aligned}
\vspace{-0.0cm}
\end{equation}
where $i=2, \ldots, N$, $j=1, \ldots, N$. 
Moreover, the partial derivatives of $\phi_{{\rm{r}},j,0}^{\circ}$ and $\theta_{{\rm{r}},j,0}^{\circ}$ with respect to $\dot{\mathbf{u}}^{\circ }$ and $\mathbf{s}_{n,l}^{\circ }$ are $\mathbf{0}$ because the angular parameters of LOS paths are not related to UE velocity and scatterer locations directly.
Similarly, the partial derivatives of the NLOS path parameters are given by
\vspace{-0.0cm}
\begin{equation}
\begin{aligned}
&\frac{\partial \phi_{n,l}^{{\rm{r}}\circ}}{\partial \mathbf{s}_{n,l}^{\circ T}}=\frac{\mathbf{c}^{{\rm{r}}\circ T}_{n,l}}{d^{\rm{r}\circ}_{n,l}\cos\theta^{{\rm{r}}\circ}_{n,l}}, &&\frac{\partial \phi_{n,l}^{{\rm{t}}\circ}}{\partial \mathbf{u}^{\circ T}}=\frac{\mathbf{c}^{{\rm{t}}\circ T}_{n,l}}{d^{\rm{t}\circ}_{n,l}\cos\theta^{{\rm{t}}\circ}_{n,l}}, &&\frac{\partial \phi_{n,l}^{{\rm{t}}\circ}}{\partial \mathbf{s}_{n,l}^{\circ T}}=-\frac{\mathbf{c}^{{\rm{t}}\circ T}_{n,l}}{d^{\rm{t}\circ}_{n,l}\cos\theta^{{\rm{t}}\circ}_{n,l}},\\[-2pt]
&\frac{\partial \theta_{n,l}^{{\rm{r}}\circ}}{\partial \mathbf{s}_{n,l}^{\circ T}}=\frac{\mathbf{v}^{{\rm{r}}\circ T}_{n,l}}{d^{\rm{r}\circ}_{n,l}}, && \frac{\partial \theta_{n,l}^{{\rm{t}}\circ}}{\partial \mathbf{u}^{\circ T}}=\frac{\mathbf{v}^{{\rm{t}}\circ T}_{n,l}}{d^{\rm{t}\circ}_{n,l}}, &&\frac{\partial \theta_{n,l}^{{\rm{t}}\circ}}{\partial \mathbf{s}_{n,l}^{\circ T}}=-\frac{\mathbf{v}^{{\rm{t}}\circ T}_{n,l}}{d^{\rm{t}\circ}_{n,l}},
\end{aligned}
\vspace{-0.0cm}
\end{equation}
where $n = 1, \ldots, N$, and $l = 1, \ldots, L_{n}$.
The partial derivatives of AOAs, e.g., $\phi_{n,l}^{{\rm{r}}\circ}$ and $\theta_{n,l}^{{\rm{r}}\circ}$, with respect to ${\mathbf{u}}^{\circ}$ and $\dot{\mathbf{u}}^{\circ}$ are $\mathbf{0}$.
Furthermore, it is also proved that $\frac{\partial \phi_{n,l}^{{\rm{t}}\circ}}{\partial \dot{\mathbf{u}}^{\circ T}}=\mathbf{0}$, and $\frac{\partial \theta_{n,l}^{{\rm{t}}\circ}}{\partial \dot{\mathbf{u}}^{\circ T}}=\mathbf{0}$.

\section{}\label{sec:cov-wls-1}
In this section, we first prove that the proposed WLS estimator in Sec. \ref{sec:wls-1-stage} is asymptotically unbiased, then we derive the covariance matrix of $\mathbf{x}$, and finally we prove that the CRLB given in Appendix \ref{sec:crlb-wls-1} can be approached by the estimator under low noise levels.

The estimation error of $\mathbf{x}$ is given by $\Delta \mathbf{x}=\mathbf{x}-\mathbf{x}^{\circ}$, 
where $\mathbf{x}^{\circ}$ is the true value of $\mathbf{x}$.
According to \eqref{eq:wls-1-error-vector} and \eqref{eq:wls-1-answer}, we derive $\Delta \mathbf{x}=\left({\mathbf{G}}^{T} \mathbf{W} {\mathbf{G}}\right)^{-1} {\mathbf{G}}^{T} \mathbf{W e}$.
Then, the statistical expectation of $\Delta \mathbf{x}$ is given by
$\mathbb{E}\{\Delta \mathbf{x}\}=\left({\mathbf{G}}^{T} \mathbf{W} {\mathbf{G}}\right)^{-1} {\mathbf{G}}^{T} \mathbf{W} \mathbb{E}\{\mathbf{e}\}$. 
Based on the linear approximation in \eqref{eq:wls-1-error-vector-approx}, we obtain $\mathbb{E}\{\mathbf{e}\}=\mathbf{0}$ under low noise levels, given that the noise term $\Delta \mathbf{m}$ is modeled as a zero mean Gaussian vector.
Consequently, we derive that $\mathbb{E}\{\Delta \mathbf{x}\}=\mathbf{0}$ and $\mathbb{E}\{\mathbf{x}\} = \mathbf{x}^{\circ}$.
Therefore, the estimator in Sec. \ref{sec:wls-1-stage} is asymptotically unbiased with low measurement noise.

Then, we derive the covariance matrix of $\mathbf{x}$. The covariance matrix is calculated by $\text{cov}(\mathbf{x}) = \mathbb{E}\left[\left(\mathbf{x} - \mathbb{E}\left(\mathbf{x}\right)\right) \left(\mathbf{x} - \mathbb{E}\left(\mathbf{x}\right)\right)^T\right]$, where $\left(\mathbf{x} - \mathbb{E}\left(\mathbf{x}\right)\right) = \left({\mathbf{G}}^{T} \mathbf{W} {\mathbf{G}}\right)^{-1} {\mathbf{G}}^{T} \mathbf{W} {\mathbf{e}}$. Thus, we can derive $\text{cov}(\mathbf{x}) = {\mathbf{G}}^{T} \mathbf{W} {\mathbf{G}}$. With low measurement noise, we have $\text{cov}(\mathbf{x}) \approx {\mathbf{G}}^{\circ T} \mathbf{W} {\mathbf{G}^\circ}$. According to \eqref{eq:wls-1-weighting-matrix}, the covariance matrix of $\mathbf{x}$ can be derived, given by $\text{cov}(\mathbf{x}) \approx\left(\left(\mathbf{B}^{-1} \mathbf{G}^\circ\right)^{T} \mathbf{Q}^{-1} \mathbf{B}^{-1} \mathbf{G}^\circ\right)^{-1}$.

Finally, we prove that $\mbox{cov}(\mathbf{x})\approx\mbox{CRLB}(\mathbf{x}^{\circ})$ under low noise levels.
Thus, we should prove that $(\left(\mathbf{B}^{-1} \mathbf{G}^\circ\right)^{T} \! \mathbf{Q}^{-1} \mathbf{B}^{-1} \mathbf{G}^\circ)^{-1} \!=\! \left(\mathbf{D}^{\circ T} \mathbf{Q}^{-1} \mathbf{D}^\circ\right)^{-1}$. 
This holds when $\mathbf{B}^{-1}\mathbf{G}^\circ \!=\! \mathbf{D}^\circ$, which if and only if $\mathbf{BD}^\circ = \mathbf{G}^\circ$.
Through direct matrix multiplications, we can finish the proof based on the identities that have been proved and used in \cite{yang2021model}.

\end{appendices}

\bibliographystyle{IEEEtran}
\bibliography{ref}{}

\end{document}